\documentclass[authoryear,preprint,review,12pt]{article}

\usepackage{mathtools,verbatim,amsmath}

\usepackage{algorithm}
\usepackage{algpseudocode}
\mathtoolsset{showonlyrefs}
\usepackage[dvipsnames]{xcolor} 
\usepackage{natbib}
\usepackage{tikz}
\usepackage{pgfplotstable}
\usepackage{enumerate}
\usepackage[colorlinks=true,linkcolor=orange,citecolor=orange]{hyperref}
\usepackage{setspace}                  
\usepackage[left=1.5in,right=1.5in,top=1.5in,bottom=1.5in]{geometry}
\usepackage{amssymb}    
\usepackage{amsmath}
\numberwithin{equation}{section}
\numberwithin{table}{section}
\numberwithin{figure}{section}
\usepackage{booktabs} 
\usepackage{multirow}
\usepackage{dcolumn} 

\newcolumntype{d}[1]{D{.}{.}{#1}}

\def\tb{\mathbf{t}}

\def\vb{\mathbf{v}}
\def\xb{\mathbf{x}}

\def\zb{\mathbf{z}}

\def\epsb{\boldsymbol{\varepsilon}}
\def\E{{\textrm E}}

\def\Xb{{\mathbf X}}

\def\Var{{\rm Var}}

\definecolor{darkolive}{HTML}{556B2F}

\pgfplotsset{compat = 1.9}
\usetikzlibrary{calc}

\begin{document}




\title{Randomized Principal Component Ensembles for High-Dimensional Calibration
    \footnote{Caren Hasler, Department of Psychology,
    Psychological Methods, Evaluation and Statistics,
    University of Zurich, Switzerland. Arnaud Tripet and Yves Till\'e, Institute of Statistics,
    University of Neuch\^atel, Switzerland
    }
    } 
\maketitle

\author{}


\begin{abstract}
Calibration is a widely used method in survey sampling to adjust weights so that estimated totals of some chosen calibration variables match known population totals or totals obtained from other sources. When a large number of auxiliary variables are included as calibration variables, the variance of the total estimator can increase, and the calibration weights can become highly dispersed. To address these issues, we propose an ensemble method based on a principal component decomposition of the auxiliary variables. We repeatedly select sets of principal components without replacement and with unequal selection probabilities. Calibration is performed on each selected set of principal components, and the resulting weights are aggregated to obtain a final set of weights. With our proposed method, it is possible to calibrate exactly for some of the main auxiliary variables, while relaxing the calibration constraints for the remaining variables. It yields a total estimator whose variance does not explode when new auxiliary variables are added while producing weights with low dispersion. Finally, our proposed method allows us to obtain a single weighting system that can be applied to several variables of interest of a survey. An estimator of the variance of the total estimator is also proposed. We evaluate the proposed total estimator and its variance estimator using a simulation study on real survey data from the Swiss Survey on Income and Living Conditions and on synthetic data. The results show that the proposed solution significantly reduces the weight variability and the variance of the total estimator compared with competing total estimators for some variables of interest.
\end{abstract}



\noindent\textbf{Keywords:}

\noindent estimation; high dimension; selection of calibration variables; weighting; weight system







\section{Introduction}\label{sec:introduction}

Calibration, as introduced by \citet{dev:sar:92}, is a widely used method in survey sampling that adjusts the sampling weights so that the estimated totals of auxiliary variables match known population totals. The estimator produced by these adjusted weights is referred to as the calibrated estimator. The main objectives of calibration are twofold: (1) to ensure consistency between survey estimated totals and known totals or totals derived from external sources, and (2) to reduce the variance of the total estimator compared to the Horvitz-Thompson estimator \citep{hor:tho:52}.
	
With the increasing availability of large-scale datasets, calibration is often performed on a large number of auxiliary variables, an approach referred to as high-dimensional calibration. In this context, however, calibration faces two major issues. First, the reduction of the variance of the total estimator can no longer be achieved. Indeed, as noted by \citet{sil:ski:97}, the Mean Squared Error (MSE) of the calibrated estimator initially decreases as more auxiliary variables are added. However, beyond a certain point, it begins to increase. Second, as highlighted by \citet{haz:bea:17:weights:review}, a large number of auxiliary variables often leads to highly dispersed calibration weights. This can result in unstable estimates for variables that are weakly correlated with the calibration variables.

Several strategies have been proposed to address these two issues. One common approach consists of selecting a subset of auxiliary variables that minimizes the estimated MSE. This can be done through best possible subset selection or forward selection procedures \citep{sil:ski:97, cha:gog:22}, by discarding some calibration constraints \citep{ban:rat:maj:92}, or by assessing the contribution of each auxiliary variable via the Shapley decomposition and retaining only the most impactful variables \citep{gua:cec:22}.
	
An alternative to variable selection is to relax some calibration constraints. This leads to so-called soft calibration. In this idea, \citet{gug:til:10} use soft calibration with mixed models. Other methods incorporate penalization, such as ridge regression approaches proposed by \citet{bea:boc:08}, \citet{rao:sin:97}, and extended by \citet{mon:ran:09}. \citet{bur:mun:rup:19} propose a generalized calibration method that remains applicable even in high-dimensional settings using soft calibration and box-constraints. \citet{wil:sav:24} propose range-restricted soft calibration independent of any particular variable of interest. Methods based on empirical likelihood have also been suggested to obtain range-restricted weights while relaxing benchmark constraints \citep{che:sit:wu:02}, or by explicitly incorporating constraint deviations into the objective function \citep{fet:gen:per:05}.

Another approach, suggested by \citet{wu:sit:01}, uses model calibration, where the calibration variables are the predicted values of the variables of interest obtained from a working model. Finally, some authors propose reducing the dimension of the calibration problem through principal component analysis \citep{car:gog:she:17}. In this case, calibration is performed not on the raw auxiliary variables but on a reduced number of principal components.
	
In this article, we propose a new calibration approach specifically designed for settings involving a large number of auxiliary variables. We propose an ensemble method based on a principal component decomposition of the auxiliary variables. We repeatedly select sets of principal components without replacement and with unequal selection probabilities, with higher probabilities assigned to components explaining a larger proportion of the variability in the auxiliary variables. For each selected set, we perform calibration using the corresponding principal components as calibration variables. The resulting weights are then aggregated to obtain a final set of calibration weights. Our approach offers several advantages. First, it reduces the instability of calibration weights generally observed in high-dimensional settings. Second, it reduces the risk of increased variance often associated with the inclusion of many auxiliary variables. Finally, as calibration is performed independently of any variable of interest, the resulting weights can be used for different estimation purposes or different variables of interest.

The article is organized as follows: in Section~\ref{sec:method} we introduce the notation and review the different methods on which our approach is based. In Section~\ref{sec:ensemble:pca} we present the proposed calibration method in detail. This section also contains a discussion on the consistency of the proposed estimator and introduces a design-based estimator of its variance. In Section~\ref{sec:simulation}, we evaluate the empirical performance of the proposed estimator and its variance estimator through Monte Carlo simulations based on real data. In Section~\ref{sec:disc}, we conclude with a discussion of the main findings. Appendix~\ref{sec:sim:SILC:robustness} assesses the robustness of the proposed method to the choice of its underlying parameters. Finally, Appendix~\ref{sec:sim:synth:data} investigates, through Monte Carlo simulations, the finite-sample performance of competing estimators for varying ratios of the sample size to the number of auxiliary variables.

\section{Framework}\label{sec:method}

\subsection{Notation}
We consider a finite population $U$ of size $N$, and denote by $k$ a generic unit with $k \in \left\{1, \ldots, N\right\}$. Let $S$ be a random sample of size $n$ selected from $U$ with a sampling design that assigns to unit $k$ an inclusion probability of appearing in the sample $\pi_k = \Pr(k \in S)$.
We assume that $\pi_k > 0$ for all $k \in U$ and denote by $d_k = 1/\pi_k$ the design weight of unit $k$. Let $y_k$ be the value of a variable of interest $y$ for unit $k \in U$. Value $y_k$ is only observed for units contained in the sample. The aim is to estimate the population total $$t_{y} = \sum_{k \in U} y_{k}$$ of variable of interest $y$. With no additional information, the total $t_y$ can be estimated by the \textit{expansion estimator} or Horvitz-Thompson (HT) estimator \citep{hor:tho:52}
\begin{align}
\label{HT}
\widehat{t}_{yd} &= \sum_{k \in S} d_k y_k.
\end{align}
The HT estimator is design-unbiased, meaning unbiased under the sampling design, provided that $\pi_k > 0$ for all $ k \in U$. We assume that the variable of interest is univariate for simplicity. However, most surveys aim to collect information on several variables of interest. The methodology presented in this article extends naturally to that case.

\subsection{Calibration}\label{sec:cal}

Consider a vector $\xb_k = \left( x_{k1}, x_{k2}, \ldots, x_{kq} \right)^\top$ of $q$ auxiliary variables for each unit $k \in U$. As commonly done in standard calibration methods, the auxiliary variables are supposed to be either continuous or binary. Multicategory categorical auxiliary variables can be accommodated by recoding them into a set of binary indicator variables prior to applying the proposed procedure. Let $\mathbf{X} \in \mathbb{R}^{N \times q}$ be the data matrix obtained by stacking the $\xb_k$, $k \in U$. That is, each row corresponds to a unit and each column to a variable. Without loss of generality, we will suppose in what follows that the columns of $\mathbf{X}$ are centered and scaled (they have mean 0 and variance 1). Denote by $\tb_\xb = \sum_{k \in U} \xb_k$ the population total of these auxiliary variables. Suppose that $\xb_k$ is available for all population units $k \in U$. We will relax this assumption in Section~\ref{sec:ensemble:pca}, and suppose instead that $\xb_k$ is available only for sampled units $k \in S$, while the population total $\tb_\xb$ remains known.

Calibration, first introduced by \cite{dev:sar:92}, allows the auxiliary information to be efficiently used to improve the HT estimator. It consists in modifying the initial design weights $d_k = 1/\pi_k$ into new weights $w_k$ that are as close as possible, in an average sense for a given metric, to the initial design weights, while satisfying the calibration equation
\begin{align}
\tb_\xb = \sum_{k \in S} w_k \xb_k.
\end{align}
This constraint ensures that the weighted sum of the auxiliary variables in the sample matches their known population totals, thereby increasing the coherence and efficiency of the estimator.
	
The weights can be written as $w_k = d_k g_k = g_k/\pi_k$, where $g_k$ solves
\begin{align}
\sum_{k \in U} \xb_k = \sum_{k \in S} \frac{g_k}{\pi_k} \xb_k.
\end{align}	
A common and convenient choice for the calibration distance is the chi-squared distance, defined as
\begin{align}
\Phi_S(\mathbf{w}) = \sum_{k \in S} \frac{(w_k - d_k)^2}{q_k d_k},
\end{align}
where $1/q_k$ is an individual scaling factor associated with each unit $k \in S$, allowing more or less importance to be assigned to certain units. When there is no reason to privilege any unit, the weights are set to $q_k = 1$ for all $k$ \citep{dev:sar:92}. Given this criterion, the calibration weights $w_k = g_k/\pi_k = d_k g_k$ are defined as the solution to the minimization problem
\begin{align}
\mathbf{w} = \arg\min_{\mathbf{w}} \Phi_S(\mathbf{w}),
\end{align}
subject to the calibration equation above. 

While calibration may introduce a small bias compared to the Horvitz– Thompson estimator, it often leads to a significant reduction in variance, particularly when the auxiliary variables $\xb$ are strongly correlated with the variable of interest $y$. This tradeoff typically results in improved mean squared error (MSE). However, when the number of auxiliary variables becomes large, calibration may lead to instability. Indeed, weights can become extremely dispersed and the variance of the estimator may increase instead of decreasing.
	
\subsection{Principal Component Analysis}\label{sec:pca}

Principal Component Analysis (PCA) is a dimensionality reduction method introduced by \citet{pea:1901} and later extended by \citet{hot:36}. It is mainly used to reduce the dimension of a dataset while preserving as much of its variability as possible. It transforms a dataset with a large set of correlated variables into a new set of uncorrelated variables called principal components, so that a smaller number of variables captures the greater part possible of the variability in the initial variables.

Let $\mathbf{X} \in \mathbb{R}^{N \times q}$ be the data matrix, where each row corresponds to a unit and each column to a variable. The goal of PCA is to find an orthonormal basis $\mathbf{V} = (\mathbf{v}_1, \dots, \mathbf{v}_q)$ such that the first few of the transformed variables, called principal components, in the columns of $\mathbf{Z} = \mathbf{X} \mathbf{V}$ capture the largest portion possible of the variance in the initial variables $\mathbf{X}$. The vectors $\mathbf{v}_j$ are solutions to the eigenvalue problem
\begin{align}
\mathbf{\Sigma} \mathbf{v}_j &= \lambda_j \mathbf{v}_j, \quad j = 1, \dots, q,
\end{align}
where $\mathbf{\Sigma}$ is the population covariance matrix of $\mathbf{X}$
\begin{align}
\mathbf{\Sigma} &= \frac{1}{N} \sum_{k=1}^N (\mathbf{x}_k - \bar{\mathbf{x}})(\mathbf{x}_k - \bar{\mathbf{x}})^\top.
\end{align}
The eigenvalues $\lambda_j$ represent the portion of the variance in the initial data matrix explained by principal component $j$ defined as
\begin{align}
\mathbf{Z}_j = \mathbf{X} \mathbf{v}_j.
\end{align}
The first $c$ principal components represent a projection of the initial data onto a lower-dimensional space. This transformation achieves a dimensionality reduction in the sense that it retains the maximum possible variance in a small number of components.

%
%

\subsection{Ensemble Methods}\label{sec:ensemble}

Ensemble methods are learning algorithms that construct a collection of estimators, predictors, or classifiers and combine their outputs to obtain a final result, see, for instance, \cite{Dietterich2000}. For example, let $\widehat{\theta}^{(b)}$, $b=1,\ldots,B$, denote a collection of estimators of a parameter $\theta$, obtained by repeating a random procedure $B$ times. A simple ensemble method consists of aggregating these estimators by taking their average. The resulting estimator is
\begin{align}
\widehat{\theta}_{\mathrm{e}} = 
\frac{1}{B}\sum_{b=1}^{B}\widehat{\theta}^{(b)}.
\end{align}
A prominent example of an ensemble method is Bootstrap Aggregating, or Bagging, introduced by \citet{bre:96}. By combining multiple estimators, aggregation can improve stability and reduce variance, as is well established for Bagging \citep{bul:yu:02}.

\subsection{Calibration on Principal Components of \cite{car:gog:she:17}}\label{sec:cardot}

Our proposed method is closely related to the calibration on principal components proposed by \cite{car:gog:she:17}, which is why we briefly present this approach here. The first step of their method is to compute the principal components of the centered and scaled matrix of auxiliary variables $\mathbf{X}$. We obtain the principal components $\mathbf{Z}_j = \mathbf{X} \mathbf{v}_j$ and eigenvalues $\lambda_j$, $j = 1, \ldots, q$, as described in Section~\ref{sec:pca}. Let $\zb_k$ denote the values of the $q$ principal components for unit $k$. Calibration is then applied to the first $c$ principal components, where $c$ is a parameter to be determined. Let $\zb_{k}^{(1:c)}$ denote the values of the first $c$ principal components for unit $k$. The calibration weights $w_k$, $k \in S$, are computed by solving the calibration equation
\begin{equation}
\sum_{k \in S} w_k \zb_{k}^{(1:c)} = \sum_{k \in U} \zb_{k}^{(1:c)},
\end{equation}
using a distance function such as the chi-square distance, see Section~\ref{sec:cal}.

The method described above assumes that the scaled matrix of auxiliary variables $\mathbf{X}$ is known for all population units. \cite{car:gog:she:17} propose an alternative approach that can be applied when the auxiliary variables are observed only for the sampled units, while their population totals are known. This approach consists of calibration on estimated principal components. We describe it below. Let $\widehat{\mathbf{\Sigma}}$ denote the estimated population covariance matrix of $\mathbf{X}$, defined by
\begin{align}
\widehat{\mathbf{\Sigma}} &= \frac{1}{\widehat{N}} \sum_{k \in S} d_k (\mathbf{x}_k - \widehat{\bar{\mathbf{x}}})(\mathbf{x}_k - \widehat{\bar{\mathbf{x}}})^\top,
\end{align}
where $\widehat{N} = \sum_{k \in S} d_k$ and $\widehat{\bar{\mathbf{x}}} = \widehat{N}^{-1}\sum_{k \in S}d_k \xb_k$. The design-based estimators $\widehat{\lambda}_j$ and $\widehat{v}_j$ of
$\lambda_j$ and $\vb_j$, respectively, for $j=1,\ldots,p$ are obtained as the solutions to the eigenvalue problem 
$\widehat{\mathbf{\Sigma}}\widehat{v}_j = \widehat{\lambda}_j \widehat{v}_j.$ See \cite{car:10} for results concerning the asymptotic design unbiasedness of these estimators. The estimated principal components are then given by $\widehat{\mathbf{Z}}_j = \mathbf{X} \widehat{\mathbf{v}}_j$. Let $\widehat{\zb}_k$ denote the values of the $q$ estimated principal components for sample unit $k$. Although these values can be computed only for sample units $k \in S$, their population totals can be obtained from the known population totals of the auxiliary variables. In particular, the population total of the $j$-th estimated principal component is
\begin{align}
  \sum_{k \in U} \widehat{z}_{kj} = \sum_{k \in U} \xb_k \widehat{v}_j,
\end{align}
where $\widehat{z}_{kj}$ denotes the value taken by the $j$-th estimated principal component for unit $k$. Thus, while the values $\widehat{z}_{kj}$ are available only for sample units, their population totals are known. Let $\widehat{\zb}_{k}^{(1:c)}$ denote the values of the first $c$ estimated principal components for unit $k$. Calibration on these estimated principal components is then defined by the calibration equation
\begin{equation}
\sum_{k \in S} w_k \widehat{\zb}_{k}^{(1:c)} = \sum_{k \in U} \widehat{\zb}_{k}^{(1:c)} = \sum_{k \in U} \xb_k \left[\widehat{v}_1 \quad \widehat{v}_2 \quad\ldots\quad \widehat{v}_c \right].
\end{equation}
Further details are given in \cite{car:gog:she:17}.

\section{Randomized Principal Component Ensembles}\label{sec:ensemble:pca}

When the number of auxiliary variables is large, traditional calibration methods can lead to highly dispersed weights and increased variance of the total estimator. To address these issues, we propose an ensemble method based on a principal component decomposition of the auxiliary variables. We repeatedly select sets of principal components without replacement and with unequal selection probabilities. For each selected set, we perform calibration using the corresponding principal components as calibration variables. The resulting weights are then aggregated to obtain a final set of weights.

In the remainder of this section, we present the proposed method and discuss the choice of its parameters, as well as its advantages and limitations. We show that the resulting calibration estimator can be expressed as a model-assisted estimator and discuss the consistency and the variance estimation of the proposed estimator. Finally, we discuss calibration in high-dimensional settings, where the number of auxiliary variables can exceed the sample size.

\subsection{Calibration via Randomized Principal Component Ensembles}\label{subsec:alt2}

In order to obtain final weights, we start by computing the principal components of the centered and scaled matrix of auxiliary variables $\mathbf{X}$. We obtain the principal components $\mathbf{Z}_j = \mathbf{X} \mathbf{v}_j$ and eigenvalues $\lambda_j$, $j = 1, \ldots, q$, as described in Section~\ref{sec:pca}. Denote by $\zb_k$ the values taken by the $q$ principal components for unit $k$. Then, we construct an ensemble by independently repeating the randomization procedure a large number of times $B$ as follows. At each iteration $b = 1, \dots, B$, we select a subset of principal components of size $c$ from among the $q$ principal components without replacement and with unequal probabilities proportional to $\lambda_j ^\alpha$. Parameter $c$ sets the number of components selected at each iteration, $\alpha$ controls the contrast in the inclusion probabilities of the components. In Section~\ref{subsec:para}, we further discuss these quantities and suggest guidelines for their choice. Let $\zb_{k}^{(b)}$ denote the values taken by the selected components for unit $k$. Calibration weights $w_k^{(b)}$, $k \in S$, are computed by solving the calibration equation
\begin{equation}
\sum_{k \in S} w_k^{(b)} \zb_{k}^{(b)} = \sum_{k \in U} \zb_{k}^{(b)},
\end{equation}
using a distance function such as the chi-square distance, see Section~\ref{sec:cal}. At the end of the $B$ iterations, we have $B$ systems of weights $w_k^{(b)}$, $b = 1, \ldots, B$. The final weights are obtained by averaging these $B$ systems of weights. That is, 
\begin{equation}
\widehat{w}_k = \frac{1}{B} \sum_{b=1}^B w_k^{(b)}.
\end{equation}
The full procedure is detailed in Algorithm~\ref{algoBagPCA}. For any variable of interest $y$, the final estimator is defined as
\begin{equation}
\widehat{t}_{ep} = \sum_{k \in S} \widehat{w}_k y_k.
\end{equation}
Subscripts $e$ and $p$ refer to ensemble and PCA, respectively. Estimator $\widehat{t}_{ep}$ can alternatively be written
\begin{align}\label{eqn:estimator}
\widehat{t}_{ep}  = \frac{1}{B} \sum_{b=1}^B \widehat{t}_{\text{cal}}^{(b)},
\end{align}
where
\begin{align}
\widehat{t}_{\text{cal}}^{(b)} = \sum_{k \in S}  w_k^{(b)} y_k
\end{align}
is the calibration estimator obtained at step $b$. Constructing an ensemble based on principal components instead of calibrating on the original variables ensures better control of weight dispersion and yields a more stable estimator.

\begin{algorithm}[htb!]
	\caption{Calibration via Randomized Principal Component Ensembles} \label{algoBagPCA}
	\begin{algorithmic}[1]
		\State Input: Auxiliary variable matrix $\mathbf{X}$, inclusion probabilities $\pi_k$, number of iterations $B$, number of principal components selected at each iteration $c$, adjustment parameter $\alpha$.
		\State Compute principal components $\mathbf{Z}_j = \mathbf{X} \mathbf{v}_j$ and eigenvalues $\lambda_j$, $j = 1, \ldots, q$.
		\For{$b = 1$ to $B$}
		\State Select a subset of $c$ principal components from among the $q$ principal components without replacement and unequal probabilities proportional to $\lambda_j^\alpha$. The values taken by the selected components for unit $k$ are $\zb_{k}^{(b)}$.
		\State Compute calibration weights $w_k^{(b)}$, $k \in S$, by solving the calibration equation
\begin{equation}
\sum_{k \in S} w_k^{(b)} \zb_{k}^{(b)} = \sum_{k \in U} \zb_{k}^{(b)}.
\end{equation}
		\EndFor
		\State Compute the final weight for each unit by aggregating the $B$ weighting systems
		\begin{align}
		\widehat{w}_k = \frac{1}{B} \sum_{b=1}^{B} w_k^{(b)}, \quad k \in S.
		\end{align}
		\State Output: final weights $\widehat{w}_k$.
	\end{algorithmic}
\end{algorithm}

Our proposed estimator is hence the average of $B$ calibrated estimators. However, our proposed estimator is usually not exactly a calibrated estimator. The final weights $\widehat{w}_k$, $k \in S$, are exactly calibrated only on those principal components that are included in each and every sample of principal components. See Section~\ref{subsec:lim} for a possible approach to obtain exact calibration on some important variables. For more details on the proposed estimator, see Section~\ref{sec:model:assisted}.

\subsection{Choice of Parameters}\label{subsec:para}

The choice of the number of selected principal components $c$ remains a critical parameter in this method. A too small value of $c$ results in weights far from being calibrated to the total of the principal components and a total estimator that does not benefit from a reduced variance in case the principal components are highly correlated to the variable of interest. As the number of selected principal components $c$ increases, the weights become closer and closer to being calibrated to the total of the principal components. When $c = q$, the weights are exactly calibrated to the total of all principal components, and therefore also to that of all initial auxiliary variables. At some point, $c$ is too large and too many principal components are selected. The problems associated with high-dimensional settings resurface, such as unstable calibration weights and total estimators. Striking the right balance between dimensionality reduction and information retention helps maximize the efficiency of the method and depends on the data at hand.
	
As a rule of thumb, we suggest $c = \min(q,\sqrt{n})$, which results in a relatively small number of calibration variables compared to the sample size. However, note that the case $c=q$, which occurs when $q$ is much smaller than $n$, in particular when $q<\sqrt{n}$, yields the usual calibration estimator. This is a desirable property since, in this setting, we are no longer in a high-dimensional regime and the usual calibration estimator does not encounter the specific issues associated with high dimensionality. Therefore, in this case, we recommend using the usual calibration estimator. Alternatively, we can fix a proportion of variance of the initial auxiliary variables to be explained by the first $c$ components. For instance, we may want to select the smallest $c$ so that the first $c$ principal components explain $60\%$ of the variance in the auxiliary variables. This selection is universal with respect to the variables of interest. Or we can also choose the smallest $c$ that explains a fixed fraction of the variation in a specific variable of interest. The resulting weights depend on one specific variable of interest and may not be suitable for another variable of interest. The finite-sample performance of competing estimators in settings with varying ratios of sample size to the number of auxiliary variables estimator and varying $c$ is investigated through Monte Carlo simulations in Appendices~\ref{sec:sim:SILC:robustness} and~\ref{sec:sim:synth:data}.
	
Another key parameter to determine is the exponent applied to the eigenvalues $\alpha$. This parameter controls how contrasted the probabilities of inclusion of the principal components are. A high value of $\alpha$ yields highly contrasted inclusion probabilities. The first principal components are much more likely to be selected than the last principal components. As $\alpha$ lowers, the contrast decreases. The inclusion probability of the principal components gets closer and closer. The choice $\alpha = 0$ assigns the same inclusion probability to all principal components. When $\alpha$ is too low,  too much emphasis is put on lower-ranked principal components, that do not explain much of the variability in the auxiliary variables, and may add noise to the calibration process. As a general guideline, we recommend $\alpha=1/2$. This choice provides a compromise between assigning similar selection probabilities to all principal components and the more deterministic case in which the leading principal components are systematically selected, which correspond to the method of \cite{car:gog:she:17}. It is also motivated by the construction of the proposed procedure, since with this choice the probability of selecting a given principal component is equal to the proportion of the standard deviation in the auxiliary variables explained by that component. The finite-sample performance of the proposed estimator in settings with varying $c$ and $\alpha$ is investigated through Monte Carlo simulations in Appendix~\ref{sec:sim:SILC:robustness}. For the considered data, many couples of values for these variables yield comparable results, with the best values obtained for values of $\alpha$ between 0 and 1.

\subsection{Advantages and Limitations}\label{subsec:lim}

The method we propose has several advantages. First, it is independent of any variable of interest. The resulting weights can therefore be applied simultaneously to several variables of interest. It is also the case of the method of \cite{car:gog:she:17}. Second, for variables of interest strongly linearly related to the principal components that are frequently selected in the iterations, the variance of the total estimator that we propose is lower than that of the HT estimator. Third, the obtained weights are very stable and close to the initial design weights. 
	
However, there are also some limitations. The calibration weights obtained with our method are not exactly calibrated, unless we decide to impose exact calibration on a set of important auxiliary variables, as presented below. In addition, if the linear relationship between a variable of interest and the first few principal components, those the most often selected in the bootstrap samples, is weak, then the performance of our total estimator may be worse than that of HT estimator for this particular variable of interest. Indeed, our estimator is biased and may not be more efficient than the HT for that variable of interest.
	
The closest competitor to our method is the method of \cite{car:gog:she:17}. With their method, the weights are exactly calibrated on the first $c$ principal components. With our method, the weights are not exactly calibrated on any principal components. Therefore, the total estimator of \cite{car:gog:she:17} is more efficient than ours for variables of interest for which a large proportion of variance is explained by the first $c$ principal components. Our total estimator is more efficient when the $q-c$ other principal components still explain a substantial portion of the variance in the variable of interest, in addition to the variance explained by the first $c$ principal components. An advantage of our method over the method of \cite{car:gog:she:17} is that the weights obtained with our method are more stable than those obtained with the method of \cite{car:gog:she:17} for a same number of principal components $c$ considered.
	
As mentioned above, the calibration weights obtained with our method are not exactly calibrated. When there is a need to obtain weights exactly calibrated on some important auxiliary variables, our method can be adapted using a simple procedure described in \cite{car:gog:she:17} and detailed in this paragraph. Suppose that we want to calibrate the weights exactly on $c_1$ auxiliary variables, where $c_1 < c < q$. We compute the residuals of the regression of the remaining $q - c_1$ auxiliary variables on these important auxiliary variables. We hence create $q - c_1$ variables of residuals. We apply PCA to these $q - c_1$ variables of residuals and obtain $q - c_1$ principal components. These principal components are orthogonal to one another and to the subspace generated by the $c_1$ important auxiliary variables. Then, at each iteration, we select all $c_1$ important auxiliary variables and a sample of $c - c_1$ principal components.

\subsection{Our Proposed Estimator as a Model-Assisted Estimator}\label{sec:model:assisted}

In this section, we show that our proposed estimator can be written as a model-assisted estimator. Our estimator is
\begin{align}\label{eqn:estimator2}
\widehat{t}_{ep} = \sum_{k \in S} \frac{g_k y_k}{\pi_k} = \frac{1}{B} \sum_{b=1}^B \widehat{t}_{\text{cal}}^{(b)},
\end{align}
where
\begin{align}
\widehat{t}_{\text{cal}}^{(b)} = \sum_{k \in S} \frac{g_k^{(b)}y_k}{\pi_k},
\end{align}
with $g_k^{(b)} $ the solution to the calibration equation
\begin{align}
\sum_{k \in U} \mathbf{z}_k^{(b)} = \sum_{k \in S} \frac{g_k^{(b)}\mathbf{z}_k^{(b)}}{\pi_k}.
\end{align}
With the chi-squared distance
\begin{align}
\Phi_S(\mathbf{w}) = \sum_{k \in S} \frac{(w_k - d_k)^2}{q_k d_k}
\end{align}
and the choice $q_k = 1$, the calibration weights $w_k^{(b)} = g_k^{(b)}/\pi_k = d_k g_k^{(b)}$ are
\begin{align}
\mathbf{w}^{(b)} = \arg\min_\mathbf{w} \Phi_S(\mathbf{w})
\end{align}
subject to the calibration equation above. The solution is
\begin{align}
g_k^{(b)} = 1 + \mathbf{z}_k^{(b)\top} \left( \sum_{k \in S}  \frac{\mathbf{z}_k^{(b)} \mathbf{z}_k^{(b)\top}}{\pi_k}  \right)^{-1}
\left( \sum_{k \in U} \mathbf{z}_k^{(b)} - \sum_{k \in S}  \frac{\mathbf{z}_k^{(b)}}{\pi_k}  \right),
\end{align}
see \cite{dev:sar:92}, Equation (1.3). The calibrated estimator $\widehat{t}_{\text{cal}}^{(b)}$ can be written as
\begin{align}
\widehat{t}_{\text{cal}}^{(b)} =
\sum_{k \in S} \frac{y_k}{\pi_k}
+ \sum_{k \in S} \frac{\mathbf{z}_k^{(b)\top}}{\pi_k} \widehat{\mathbf{T}}_{\mathbf{z}\pi}^{(b)-1} \mathbf{t}_{\mathbf{z}}^{(b)} y_k
- \sum_{k \in S} \frac{\mathbf{z}_k^{(b)\top}}{\pi_k} \widehat{\mathbf{T}}_{\mathbf{z}\pi}^{(b)-1}\widehat{\mathbf{t}}_{\mathbf{z}\pi}^{(b)}y_k,
\end{align}
where
\begin{align}
  \widehat{\mathbf{T}}_{\mathbf{z}\pi}^{(b)} &= \sum_{k \in S}  \frac{\mathbf{z}_k^{(b)} \mathbf{z}_k^{(b)\top} }{\pi_k}  ,\\
  \mathbf{t}_{\mathbf{z}}^{(b)} &= \sum_{k \in U} \mathbf{z}_k^{(b)},\\
  \widehat{\mathbf{t}}_{\mathbf{z}\pi}^{(b)} &= \sum_{k \in S}  \frac{\mathbf{z}_k^{(b)}}{\pi_k} .
\end{align}
Rearranging, we obtain
\begin{align}\label{eqn:tcal}
\widehat{t}_{\text{cal}}^{(b)} = \sum_{k \in U} \mathbf{z}_k^{(b)\top} \widehat{\mathbf{B}}_S^{(b)} +
\sum_{k \in S} \frac{y_k - \mathbf{z}_k^{(b)\top} \widehat{\mathbf{B}}_S^{(b)}}{\pi_k}.
\end{align}
where
\begin{align}
\widehat{\mathbf{B}}_S^{(b)} &= \left(\widehat{\mathbf{T}}_{\mathbf{z}\pi}^{(b)}\right)^{-1}\sum_{k \in S}  \frac{\mathbf{z}_k^{(b)}}{\pi_k}  y_k\\
&=\left( \sum_{k \in S}  \frac{\mathbf{z}_k^{(b)}\mathbf{z}_k^{(b)\top}}{\pi_k}   \right)^{-1}
\sum_{k \in S}  \frac{\mathbf{z}_k^{(b)}}{\pi_k}  y_k
\end{align}
Finally, using Equations~\eqref{eqn:estimator2} and \eqref{eqn:tcal}, our proposed estimator $\widehat{t}_{ep} $ can be rewritten as a model-assisted estimator as follows
\begin{align}
\widehat{t}_{ep} = \sum_{k \in U} \widehat{m}(\mathbf{z}_k) + \sum_{k \in S} \frac{y_k - \widehat{m}(\mathbf{z}_k)}{\pi_k},
\end{align}
where
\begin{align}
\widehat{m}(\mathbf{z}_k) = \frac{1}{B} \sum_{b=1}^B \mathbf{z}_k^{(b)\top} \widehat{\mathbf{B}}_S^{(b)}.
\end{align}
The predicted value $\widehat{m}(\mathbf{z}_k)$ is the average over all samples $b = 1, \ldots , B$ of the predictions $\mathbf{z}_k^{(b)\top} \widehat{\mathbf{B}}_S^{(b)}$ of the linear regressions of the variable of interest on the principal components included in these samples.

\subsection{Consistency and Variance Estimation}\label{sec:variance}

The proposed estimator is an average of calibration estimators, see Equation~\eqref{eqn:estimator}. Each ensemble member $\widehat{t}_{\text{cal}}^{(b)}$, $b=1,\ldots,B$ is a calibration estimator based on a subset of principal components. Under the conditions established by \citet{dev:sar:92} and \citet{isa:ful:82}, and more rigorously by \citet{car:gog:she:17} for calibration on principal components, $N^{-1}\widehat{t}_{\text{cal}}^{(b)}$, $b=1,\ldots,B$ is consistent for $N^{-1}t_y$, provided that the number $c$ of selected principal components remains sufficiently small relative to the sample size. Since the proposed estimator averages such calibration estimators over the ensemble, $N^{-1}\widehat{t}_{ep}$ inherits this consistency under the same conditions.

Applying the law of iterated variances, the variance of $\widehat{t}_{ep}$ can be decomposed as
\begin{align}\label{eqn:variance}
\Var\left(\widehat{t}_{ep}\right) &= V_1 + V_2,\\
    V_1
    &=
    \E_d\left[
        \Var_{\mathrm{rpc}}\left(\widehat{t}_{ep}\mid S\right)
    \right],
    \\
    V_2 &=
    \Var_d\left[
        \E_{\mathrm{rpc}}\left(\widehat{t}_{ep}\mid S\right)
    \right],
    \end{align}
where subscripts $d$ and rpc denote expectations and variances under the sampling design and the randomized principal component selection scheme, respectively.

The first term, $V_1$, represents the additional variability induced by the random selection of the components. The second term, $V_2$, represents the sampling variability of the estimator after averaging over the randomized selection scheme. When the number of ensemble members $B$ is sufficiently large, we expect the contribution of $V_1$ to be small relative to that of $V_2$. The results of our simulation study presented in Section~\ref{sec:sim:var} support this approximation. For small values of $B$, the variability induced by the randomized component selection may have a larger contribution to the overall variance. As $B$ increases, averaging over a larger number of randomized selections reduces this source of variability. At the same time, the random selection and subsequent averaging can stabilize the calibration procedure by reducing its sensitivity to any particular subset of principal components. Hence, we use the approximation
    $\Var\left(\widehat{t}_{ep}\right)
    \approx
        \Var_d\left(
        \widehat{t}_{ep}\right)$,
where we used $\E_{\mathrm{rpc}}\left(\widehat{t}_{ep}\mid S\right) \approx \widehat{t}_{ep}$ because for large numbers of ensemble members $B$, the variability due to the randomization of the principal components is negligible conditional on $S$.

Our estimator $\widehat{t}_{ep}$ is an average of calibration estimator, i.e.
\begin{align}
\widehat{t}_{ep}  = \frac{1}{B} \sum_{b=1}^B \widehat{t}_{\text{cal}}^{(b)}.
\end{align} 
We apply the approach of \citet{dev:sar:92} based on residuals for estimating the variance of each calibrated estimator $\widehat{t}_{\text{cal}}^{(b)}$. We can then linearize each calibrated estimator on a subset of principal component, yielding the following approximation to the difference between the estimator and the target parameter
\[
\widehat{t}_{\mathrm{cal}}^{(b)} - t_y \approx \sum_{k\in S} \frac{e_k^{(b)}}{\pi_k},
\]
where
\[
e_k^{(b)} = y_k -
\mathbf{z}_k^{(b)\top}
\left(
\sum_{\ell\in S} w_\ell^{(b)} q_\ell \mathbf{z}_\ell^{(b)} \mathbf{z}_\ell^{(b)\top}
\right)^{-1}
\sum_{\ell\in S} w_\ell^{(b)} q_\ell \mathbf{z}_\ell^{(b)} y_\ell,
\]
and $q_k = 1$ for all $k$ if there is no reason to privilege any unit in the calibration, see Section~\ref{sec:cal}.

This result directly provides a linearization of the proposed estimator $\widehat{t}_{ep}$. Indeed, we obtain
\[
\widehat{t}_{ep} - t_y
\approx
\sum_{k\in S} \frac{e_{k,ep}}{\pi_k},
\]
where
\[
e_{k,ep}
=
\frac{1}{B}\sum_{b=1}^B e_k^{(b)}.
\]

The variance of $\widehat{t}_{ep}$ can therefore be estimated using a standard plug-in approach based on the aforementioned approximation $\Var\left(\widehat{t}_{ep}\right)
    \approx
        \Var_d\left(
        \widehat{t}_{ep}\right)$, by replacing the linearized variables in the usual design-based variance estimator with the averaged residuals $e_{k,ep}$. This gives
\begin{align}\label{eqn:var:hat}
\widehat{V}(\widehat{t}_{ep})
=
\frac{n}{n-c}\sum_{k\in S}\sum_{\ell\in S} \frac{\pi_{k\ell}-\pi_k\pi_\ell}{\pi_{k\ell}} \frac{e_{k,ep}}{\pi_k}\frac{e_{\ell,ep}}{\pi_\ell}.
\end{align}
where $\pi_{k\ell}$ denotes the second-order inclusion probability, with the convention $\pi_{kk}=\pi_k$, and factor $n/(n-c)$ corrects for the loss of degrees of freedom associated with the selection of $c$ components. The finite-sample performance of this variance estimator is investigated through Monte Carlo simulations in Section~\ref{sec:sim:var}.

We expect the proposed variance estimator to underestimate the variance for two reasons. First, it is based only on $V_2$ and does not account for the component $V_1$. As discussed above, the contribution of $V_1$ is expected to be small when $B$ is sufficiently large. Second, and potentially more importantly, the same sample observations are used both to fit the calibration models, i.e. to compute the $B$ regression coefficients, and to compute the residuals.  The resulting in-sample residuals may therefore underestimate the residuals relevant for variance estimation, leading to an underestimation of the variance. One possible remedy would be to construct the residuals using a cross-validation procedure. A detailed investigation of such corrections is beyond the scope of the present work and is left for future research.

\subsection{Calibration when the Number of Auxiliary Variables is Large Relative to the Sample Size}

In an extreme setting such as $n<q$, as considered in our simulation study in Section~\ref{sec:simulation}, the usual calibration estimator is unlikely to be applied in practice. When the number of auxiliary variables is large relative to the sample size, alternative calibration approaches may therefore be considered. For instance, ridge calibration can be used \citep{bea:boc:08}. In practice, it may also be desirable to impose range restrictions on the calibration weights, by requiring them to lie between prespecified lower and upper bounds, in order to avoid extreme weights. Such restrictions, however, do not guarantee a reduction in the variance of estimators of the variables of interest and may even lead to an increase in variance.

Ridge calibration provides a compromise between satisfying the calibration equations and controlling the calibration weights. More specifically, it allows discrepancies between the estimated and known population totals of the auxiliary variables in order to obtain a solution satisfying range restrictions on the weights. Exact calibration may nevertheless be retained for a subset of auxiliary variables considered particularly important.

Ridge calibration and the approach proposed in this paper have several similarities. Neither necessarily achieves exact calibration on the complete set of auxiliary variables, while both allow exact calibration on a subset of important auxiliary variables. Moreover, both approaches require the specification of tuning parameters: a regularization parameter for ridge calibration and the parameters $\alpha$ and $c$ for our approach. One difference is that range restrictions on the weights can be explicitly incorporated into ridge calibration, whereas our approach does not impose such restrictions per se. Nevertheless, the weights obtained with our approach are generally stable and concentrated around the design weights.  Range restrictions, as imposed in ridge calibration, could further constrain the calibration problem and potentially increase the discrepancy between the estimated and known population totals of the auxiliary variables.

Another difference concerns how departures from exact calibration arise. In our approach, the calibration equations are satisfied exactly at each individual calibration step. However, after aggregating the resulting estimators, the final weights do not, in general, satisfy these calibration equations exactly. The proposed approach may be particularly useful in the presence of strong multicollinearity among the auxiliary variables, when a small number of principal components can capture a substantial proportion of their variability. It may also be advantageous when a variable of interest is strongly related to some of the leading principal components, which are then likely to be frequently selected for use in the individual calibration steps.

Finally, in such extreme settings, the Horvitz-Thompson estimator may also constitute a serious competitor. Since it retains the original design weights, it avoids the potentially extreme weights resulting from calibration. On the other hand, it does not exploit the available auxiliary information to improve efficiency. This may represent an important limitation when a variable of interest is strongly related to some of the leading principal components of the auxiliary variables. The finite-sample performance of competing estimators in settings with varying ratios of sample size to the number of auxiliary variables estimator is investigated through Monte Carlo simulations in Appendix~\ref{sec:sim:synth:data}.

\section{Simulation Study}\label{sec:simulation}

\subsection{Data Presentation and Preparation}

\begin{figure}[htb!]
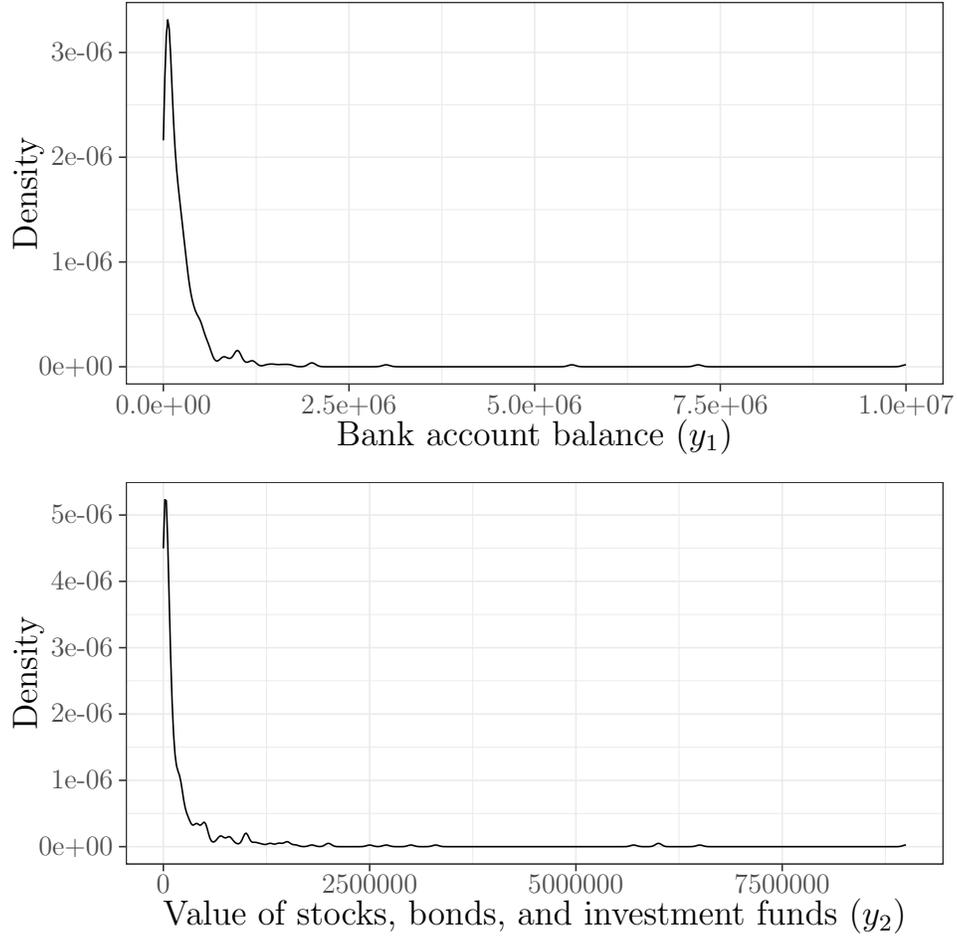

		\centering

\caption{Density estimates of Bank account balance, $y_1$ (top panel) and Value of stocks, bonds, and investment funds, $y_2$ (bottom panel) for 425 households.}\label{fig:var:interest}
\end{figure}

We conduct a simulation study on a real dataset to assess the accuracy of our proposed estimator in a practical context. We analyze multiple scenarios to ensure robustness of the presented results. We consider the Swiss Survey on Income and Living Conditions (SILC) data \citep{sfo:15:silc}. The considered dataset consists of 89 variables recorded for $N = 425$ households. Among the 89 variables, $q = 87$ serve as auxiliary variables, with 64 being binary (dummy) and 23 being continuous. The remaining 2 are the variables of interest. Among the auxiliary variables we find for instance a variable indicating whether the household is located in Zurich or not, a variable that records the number of persons in the household, and the mortgage amount of the household. The variables of interest are:
\begin{itemize}
	\item $y_1$: bank account balance and
	\item $y_2$: value of stocks, bonds, and investment funds.
\end{itemize}
Figure~\ref{fig:var:interest} shows density estimates of the variables of interest. We see that these variables of interest are highly skewed to the right with large outliers.
	
\begin{figure}[htb!]
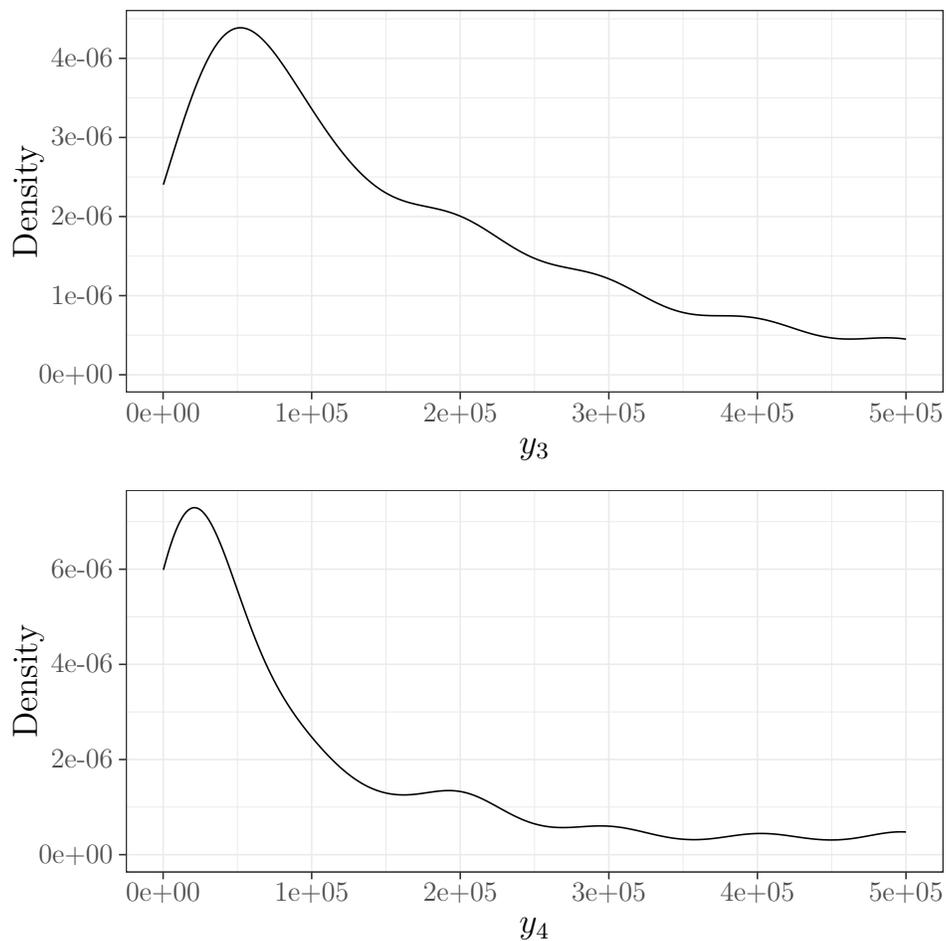

		\centering

\caption{Density estimates of bank account balance without values above 500,000, $y_3$ (top panel) and Value of stocks, bonds, and investment funds without values above 500,000, $y_4$ (bottom panel) for 425 households.}\label{fig:var:interest2}
\end{figure}

In order to test the influence of extreme values on our proposed approach, we create two additional variables of interest as follows. We create a variable of interest $y_3$ that is a copy of $y_1$, except that all values above 500,000 are replaced with random draws from the remaining distribution. We repeat the same procedure to create a copy $y_4$ of $y_2$. Figure~\ref{fig:var:interest2} shows density estimates of the created variables of interest $y_3$ and $y_4$. We see that $y_3$ and $y_4$ are still right-skewed, but much less so than $y_1$ and $y_2$. Moreover, $y_3$ and $y_4$ do not show any clear outliers.
	
Each auxiliary variable is then scaled by subtracting its mean and dividing by its standard deviation. We use the scaled auxiliary variables in what follows. We apply the PCA decomposition to the matrix $\Xb$ of scaled auxiliary variables as described in Section~\ref{sec:pca}. The obtained eigenvalues $\lambda_i$, $i \in 1, \ldots, 87$ are ranked in decreasing order $\lambda_1 = 7.14 > \lambda_2 = 5.84 > \ldots > \lambda_{87} = 6.48\cdot10^{-7}$. Their values range from approximately 0 to 7.14, with a mean value of 1 and a median value of 0.65. We choose the probabilities of inclusion of the principal components in the set of calibration variables to be proportional to the square root of the corresponding eigenvalues. We follow the recommendations given in Section~\ref{subsec:para} to set the values of the parameters $c$ and $\alpha$. We select $c = 10 \approx \min(q, \sqrt{n})$ (see below for the value of $n$) components to serve as calibration variables at each iteration. The first $10$ components explain 43\% of the variability in the auxiliary variables. We set $\alpha = 1/2$, which means that the probability that principal component $\mathbf{Z}_i$ is a calibration variable is proportional to $\lambda_i^{1/2}$. Since $\lambda_i$ represents the variance explained by the $i$th principal component, this choice makes its selection probability proportional to the standard deviation it explains.
	
Table~\ref{tab:r:squared} contains the R-squared values for linear regressions of $y_1$ to $y_4$ on all the auxiliary variables and on the first 10 principal components. Comparing the values for $y_1$ (respectively $y_2$) and those for $y_3$ (respectively $y_4$), we can see that cutting the tail of the distribution decreases by more than a half the proportion of the variance in the variables of interest explained by the linear models. Moreover, we also see an important drop in the proportion of the variance in the variables of interest explained by the model when the first 10 principal components are considered as compared to when all the auxiliary variables are considered.
	
\begin{table}[htb!]
		\centering
		\caption{R-squared values for linear regressions of $y_1$ to $y_4$ on all auxiliary variables and on the first 10 principal components.}
		\label{tab:r:squared}
		\begin{tabular}{lrr}
			\toprule
			& \textbf{All Auxiliary Variables} & \textbf{First 10 Principal Components} \\
			\midrule
			$y_1$ & 0.670 & 0.230\\
			$y_2$ & 0.637 & 0.305\\
			$y_3$ & 0.305 & 0.097\\
			$y_4$ & 0.260 & 0.069\\
			\bottomrule
		\end{tabular}
\end{table}

\subsection{Simulation Design}\label{sec:sim:dsgn}
	
We carry out 10,000 simulation runs as follows. For each simulation run, a sample of 20\% of the households is selected with simple random sampling without replacement. This corresponds to a sample size of $n = 85$. We have $n = 85 <q = 87$. In that case, the matrix involved in the computation of the linear calibration weights is singular and the sampling R package \citep{mat:til:23} computes the calibration weights using a Moore–Penrose generalized inverse. The total of the four variables of interest is computed using five different estimators. All five estimators can be written as calibration estimators of the form $\sum_{k \in S} \frac{g_k}{\pi_k}y_k$ with different choices of coefficients $g_k$. These estimators are described below.
\begin{enumerate}
	\item (CAL) The usual calibration estimator. This estimator is obtained with all auxiliary variables as calibration variables.
	\item (PCA) The calibration estimator with the first $c = 10$ principal components used as calibration variables. This is the method proposed in \cite{car:gog:she:17}.
	\item (VE) The variable ensemble calibration estimator, where $c = 10$ auxiliary variables selected at random with equal probabilities for each ensemble member. We use $B = 500$. 
	\item (PCE) The principal component ensemble calibration estimator with $c = 10$ principal components selected at random with probabilities proportional to the square root of the eigenvalues of the sample covariance matrix ($\alpha = 1/2$) as explained in the previous section. We use $B = 500$ ensemble members. We use functions ``UPMEpiktildefrompik'', ``UPMEqfromw'', and ``UPMEsfromq'' of  R package ``sampling'' \citep{mat:til:23} in order to select the samples of principal components with fixed sample size $c = 10$ and unequal probabilities.
	\item (HT) The Horvitz-Thompson estimator, which is obtained with $g_k = 1$ for all $k \in S$.
\end{enumerate}
For the first four estimators, we consider linear calibration, that is the chi-squared distance, and use function ``calib'' of R package ``sampling'' \citep{mat:til:23} to obtain the coefficients $g_k$. We include a constant calibration variable for all four estimators and set $q_k = 1$, see Section~\ref{sec:cal}.

\subsection{Measures of Comparison and Results for the Point Estimator}
	
For a generic mean estimator $\widehat{t}$, we compute the following measures of comparison.
\begin{itemize}
	\item The Monte Carlo relative Bias (RB)
	$$
	\mbox{RB} = \frac{I^{-1} \sum_{i=1}^I \widehat{t}^{(i)} - t}{t},
	$$
	where $\widehat{t}^{(i)}$ is the value of $\widehat{t}$ obtained at simulation run $i$, $i =  1, \ldots, I$, $I=10,000$ is the number of simulations, and $t$ the population total that estimator $\widehat{t}$ intends to estimate.
	\item The Monte Carlo Relative Standard Deviation (RSD)
	$$
	\mbox{RSD} = \frac{\left[(I-1)^{-1} \sum_{i=1}^I \left(\widehat{t}^{(i)}  - \overline{\widehat{t}}^{(\cdot)}\right)^2\right]^{1/2}}{t},
	$$
	where $\overline{\widehat{t}}^{(\cdot)} = I^{-1} \sum_{i=1}^I \widehat{t}^{(i)}$ is the mean value of $\widehat{t}$ over all simulation runs.
	\item Monte Carlo Relative Root Mean Square Error (RRMSE)
	$$
	\mbox{RRMSE} = \frac{\left[(I-1)^{-1} \sum_{i=1}^I \left(\widehat{t}^{(i)}  - t\right)^2\right]^{1/2}}{t}.
	$$
	\item The Monte Carlo Variance relative to the Monte Carlo Variance of the HT estimator
	$$
	\mbox{VARrHT} = \frac{(I-1)^{-1} \sum_{i=1}^I \left(\widehat{t}^{(i)}  - \overline{\widehat{t}}^{(\cdot)}\right)^2}{(I-1)^{-1} \sum_{i=1}^I \left(\widehat{t}_{HT}^{(i)}  - \overline{\widehat{t}}_{HT}^{(\cdot)}\right)^2}.
	$$
\end{itemize}
	
The results are presented in Table~\ref{tab:mc:results}. All values reported in the table are relative to the corresponding true totals and are not expressed as percentages. In this high-dimensional setting, the usual calibration estimator (CAL) performs poorly. It is highly biased, with a bias between approximately 25\% and 71\% of the corresponding true totals. It is also highly unstable, with a standard deviation between approximately six and ten times the corresponding true totals across the variables of interest. The estimator based on an ensemble of the auxiliary variables selected with equal probabilities (VE) performs even worse. In the worst case, its bias amounts to approximately six times the value of the true total. The estimator is also highly unstable with a standard deviation up to approximately 100 times the corresponding true totals.

The method of \cite{car:gog:she:17} (PCA) and our method (PCE) both provide excellent results in terms of bias with relative biases (in absolute value) smaller than 5\% of the corresponding true totals. Our proposed estimator provides the most stable estimators with a variance smaller than that of the HT estimator for all four variables of interest. For variables of interest $y_1$ and $y_2$ the variance of our proposed total estimator is approximately 74\% of that of the HT estimator. That is, there is a reduction of 26\% of the variance as compared to the HT estimator. For variables of interest $y_3$ and $y_4$, the variance of our proposed total estimator is almost the same as that of the HT estimator. The reason is that a linear model predicting the variable of interest based on the auxiliary variables and on the principal components explains a greater part of the variance in $y_1$ and $y_2$ than in $y_3$ and $y_4$, as indicated by the R-squared values in Table~\ref{tab:r:squared}. We can also note that our proposed estimator (PCE) provides better results than the estimator of \cite{car:gog:she:17} (PCA). The reason is that the linear model of the variables of interest on the first ten principal components (the components on which their estimator is calibrated) explain much less of the variability in the variables of interest than the linear models on all the auxiliary variables. That is, some of the variability in the variables of interest is explained by the remaining components. Our proposed method uses these components, the method of \cite{car:gog:she:17} does not. In a context where the first principal components explain a larger portion of the variability in the variables of interest, the method of \cite{car:gog:she:17} performs better than it does in the current context.
	
We conducted a robustness check to examine how our method behaves under different values of $\alpha$ and $c$, see Appendix. Section~\ref{sec:sim:SILC:robustness}.
	
\begin{table}[htb!]
		\centering
		\caption{Monte Carlo relative Bias (RB), Relative Standard Deviation (RSD), Relative Root Mean Square Error (RRMSE), and Variance relative to the Monte Variance of the HT estimator (VARrHT) for five estimators and four variables of interest.}
		\label{tab:mc:results}
\begin{tabular}{rl d{4.2} d{5.2} d{5.2} d{4.4} }
\toprule
$y$ & \textbf{Method} & \multicolumn{1}{r}{\textbf{RB}} &  \multicolumn{1}{r}{\textbf{RSD}} &  \multicolumn{1}{r}{\textbf{RRMSE}} &  \multicolumn{1}{r}{\textbf{VARrHT}} \\
\midrule
1 & CAL & 0.41 & 9.31 & 9.32 & 1.52e3\\
1 & PCA & 0.03 & 0.30 & 0.30 & 1.59\\
1 & VE & 2.78 & 102.02 & 102.06 & 1.83e5\\
1 & PCE & -0.03 & 0.20 & 0.21 & 0.74\\
1 & HT & 0.00 & 0.24 & 0.24 & 1.00\\
\midrule
2 & CAL & 0.71 & 10.71 & 10.73 & 1.58e3\\
2 & PCA & -0.01 & 0.26 & 0.26 & 0.94\\
2 & VE & 6.33 & 97.59 & 97.79 & 1.31e5\\
2 & PCE & -0.05 & 0.23 & 0.24 & 0.74\\
2 & HT & 0.00 & 0.27 & 0.27 & 1.00\\
\midrule
3 & CAL & 0.25 & 6.46 & 6.46 & 6.27e3\\
3 & PCA & 0.00 & 0.09 & 0.09 & 1.21\\
3 & VE & -1.69 & 38.36 & 38.40 & 2.21e5\\
3 & PCE & 0.00 & 0.08 & 0.08 & 0.98\\
3 & HT & 0.00 & 0.08 & 0.08 & 1.00\\
\midrule
4 & CAL & 0.50 & 9.85 & 9.86 & 6.47e3\\
4 & PCA & 0.02 & 0.13 & 0.14 & 1.21\\
4 & VE & 1.15 & 50.83 & 50.84 & 1.72e5\\
4 & PCE & 0.00 & 0.12 & 0.12 & 0.98\\
4 & HT & 0.00 & 0.12 & 0.12 & 1.00\\
\bottomrule
\end{tabular}
\end{table}

\subsection{Measure of Comparison and Results for the Calibration Coefficients $g_k$}
	
For each simulation run, we compute the Coefficient of Variation (CV) of the calibration coefficients $g_k$ for each of the four calibration estimators (CAL, PCA, VE, PCE) defined as the ratio of the standard deviation of the $g_k$'s to their mean.
We obtain 10,000 CVs for each estimator. Table~\ref{tab:cv} contains summary statistics of the CVs of the coefficients $g_k$ over 10,000 simulation runs for the four calibration methods. We can see that the CVs obtained with estimators CAL and VE are very high compared to the CVs of PCA and PCE. Highly dispersed coefficients may be problematic, especially if the coefficients are used to compute other parameters of interest such as totals in small domains. This illustrates how CAL and VE are inappropriate in the context of high dimension. PCA and PCE provide the smallest CV with values between 0.10 and 1.27 for PCA and between 0.07 and 0.24 for PCE. Our proposed estimator (PCE) provides the best results with CVs much smaller. This illustrates how our proposed method stabilizes the calibration coefficients $g_k$, and hence the final weights, in the context of high dimension.
	
\begin{table}
		\centering
		\caption{Summary Statistics of the CVs of the coefficients $g_k$ over 10,000 simulation runs for four methods.}
		\label{tab:cv}
		\begin{tabular}{lrrrrrr}
			\toprule
			\textbf{Method} & \textbf{Min.} & \textbf{1st Qu.} & \textbf{Median} & \textbf{Mean} & \textbf{3rd Qu.} & \textbf{Max.} \\
\midrule
CAL & 0.83 & 2.26 & 15.43 & 52.58 & 84.90 & 665.75\\
PCA & 0.10 & 0.30 & 0.37 & 0.39 & 0.47 & 1.27\\
VE & -1606.14 & 0.19 & 16.31 & 27.45 & 33.69 & 1236.65\\
PCE & 0.07 & 0.12 & 0.13 & 0.13 & 0.14 & 0.24\\
\bottomrule
		\end{tabular}
\end{table}

\subsection{Measures of Comparison and Results for the Variance Estimator}\label{sec:sim:var}

In order to compute the two terms of the variance in Equation~\eqref{eqn:variance}, for each selected sample $S^{(i)}$, $i = 1, \ldots, I$, we compute $J=500$ proposed estimators $\widehat{t}_{ep}^{(i,j)}$, $j = 1, \ldots, J$. We also compute the value of the variance estimator $\widehat{V}^{(i,j)}$ described in Section~\ref{sec:variance}. Table~\ref{tab:var} contains the following measures:
\begin{itemize}
  \item The Monte Carlo variance of the proposed estimator ($\Var$)
	$$
	\Var = (I-1)^{-1} \sum_{i=1}^I \left(\widehat{t}_{ep}^{(i,1)}  - I^{-1} \sum_{i=1}^I \widehat{t}_{ep}^{(i,1)}\right)^2.
	$$
 \item The Monte Carlo estimate of $V_1$
 	$$
	V_1 = I^{-1} \sum_{i=1}^I (J-1)^{-1} \sum_{j = 1}^J \left(\widehat{t}_{ep}^{(i,j)}  - J^{-1} \sum_{j=1}^J \widehat{t}_{ep}^{(i,j)}\right)^2,
	$$
  \item The Monte Carlo estimate of $V_2$
 	$$
	V_2 = (I-1)^{-1} \sum_{i=1}^I \left\{ J^{-1} \sum_{j = 1}^{J} \widehat{t}_{ep}^{(i,j)} - I^{-1}J^{-1} \sum_{i = 1}^{I}\sum_{j=1}^J \widehat{t}_{ep}^{(i,j)}  \right\}^2 ,
	$$
\item The Monte Carlo mean of the variance estimator ($\overline{\widehat{V}}$)
    $$
        \overline{\widehat{V}} = I^{-1} \sum_{i = 1}^{I} \widehat{V}^{(i,1)},
    $$
\item  The Monte Carlo relative bias of the variance estimator (RB)
    $$
        \mbox{RB} = \left(\overline{\widehat{V}} - \Var\right)/\Var,
    $$
\item The Monte Carlo coverage rate of the 95\% confidence interval based on the normality assumption (CR). This corresponds to the proportion of simulation runs in $i = 1, \ldots, I$  for which the true total falls within $\left[ \widehat{t}^{(i,1)}_{ep} - 1.96 \sqrt{\widehat{V}^{(i,1)}};  \widehat{t}^{(i,1)}_{ep} + 1.96 \sqrt{\widehat{V}^{(i,1)}}\right]$.
\end{itemize}

\begin{table}
		\centering
		\caption{Monte Carlo variance of the proposed estimator ($\Var$), estimate of $V_1$, estimate of $V_2$, mean of the variance estimator ($\overline{\widehat{V}}$), relative bias of the variance estimator (RB), and coverage rate (CR) for four variables of interest.}
		\label{tab:var}
\begin{tabular}{rrrrrrr}
\toprule
$y$ & $\Var$ & $V_1$ & $V_2$ & $\overline{\widehat{V}}$ & RB & CR\\
\midrule
1 & 6.32e8 & 5.00e5 & 6.32e8 & 4.46e8 & -0.29 & 0.73\\
2 & 8.46e8 & 5.77e5 & 8.44e8 & 6.50e8 & -0.23 & 0.80\\
3 & 2.95e7 & 8.95e3 & 2.95e7 & 2.57e7 & -0.13 & 0.93\\
4 & 2.60e7 & 7.81e3 & 2.60e7 & 2.29e7 & -0.12 & 0.93\\
\bottomrule
\end{tabular}
\end{table}

These results show that the contribution of $V_1$ to the variance is small relative to that of $V_2$ when the number of ensemble members $B$ is sufficiently large, with $B=500$ in this simulation. Moreover, the proposed variance estimator tends to underestimate the true variance, with a negative RB ranging from $-29\%$ to $-12\%$. This underestimation contributes to low coverage rates for $y_1$ and $y_2$, equal to $73\%$ and $80\%$, respectively. In contrast, the coverage rates for $y_3$ and $y_4$ are both $93\%$, which is close to the nominal level of $95\%$. The downward bias of the variance estimator is discussed further in Section~\ref{sec:variance}.

\section{Discussion} \label{sec:disc}
	
In this paper, we introduce a new ensemble calibration approach for high-dimensional settings, based on randomized principal components calibration. Conventional calibration often fails when the number of auxiliary variables is high, leading to unstable weights and total estimators. To overcome this problem, we perform multiple calibrations using random subsets of principal components and aggregate the resulting weights. This aggregation strategy improves stability and delivers good performance for different variables of interest.
	
Our simulation results show several advantages of the proposed method. First, the final weights obtained by aggregation are considerably less dispersed than with standard calibration or calibration based on PCA alone. Second, the variance of the total estimator remains small, even in the presence of many irrelevant or redundant auxiliary variables. Moreover, the final weights are not tailored to a specific variable of interest $y$, allowing them to be reused for multiple variables, which is particularly useful for surveys with many variables of interest.

\par
\section*{Acknowledgements}

This research was partially supported by the Swiss Statistical Office. The views expressed in this article are those of the author solely and do not necessarily reflect those of the aforementioned organization.
\par

\appendix

\refstepcounter{section}
\section*{Appendix \thesection: Choice of the Number of Principal Components $c$ and Exponent $\alpha$}\label{sec:sim:SILC:robustness}

We conducted a robustness check to examine how our method behaves under different values of $c$ and $\alpha$. We run 10,000 simulations as explained earlier in Section~\ref{sec:simulation} for different values of $c$ and $\alpha$. Below are the results for $y_1$. The results for the other three variables of interest are similar.

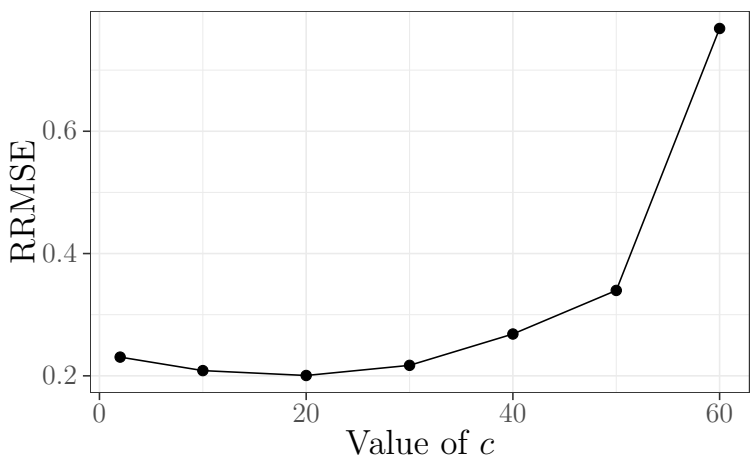
\begin{figure}[htb!]
		\centering
\begin{tikzpicture}[x=1pt,y=1pt]
\definecolor{fillColor}{RGB}{255,255,255}
\path[use as bounding box,fill=fillColor,fill opacity=0.00] (0,0) rectangle (361.35,216.81);
\begin{scope}
\path[clip] (  0.00,  0.00) rectangle (361.35,216.81);
\definecolor{drawColor}{RGB}{255,255,255}
\definecolor{fillColor}{RGB}{255,255,255}

\path[draw=drawColor,line width= 0.6pt,line join=round,line cap=round,fill=fillColor] (  0.00,  0.00) rectangle (361.35,216.81);
\end{scope}
\begin{scope}
\path[clip] ( 38.56, 31.75) rectangle (355.85,211.31);
\definecolor{fillColor}{RGB}{255,255,255}

\path[fill=fillColor] ( 38.56, 31.75) rectangle (355.85,211.31);
\definecolor{drawColor}{gray}{0.92}

\path[draw=drawColor,line width= 0.3pt,line join=round] ( 38.56, 64.28) --
	(355.85, 64.28);

\path[draw=drawColor,line width= 0.3pt,line join=round] ( 38.56,120.07) --
	(355.85,120.07);

\path[draw=drawColor,line width= 0.3pt,line join=round] ( 38.56,175.85) --
	(355.85,175.85);

\path[draw=drawColor,line width= 0.3pt,line join=round] ( 52.98, 31.75) --
	( 52.98,211.31);

\path[draw=drawColor,line width= 0.3pt,line join=round] (117.08, 31.75) --
	(117.08,211.31);

\path[draw=drawColor,line width= 0.3pt,line join=round] (181.18, 31.75) --
	(181.18,211.31);

\path[draw=drawColor,line width= 0.3pt,line join=round] (245.28, 31.75) --
	(245.28,211.31);

\path[draw=drawColor,line width= 0.3pt,line join=round] (309.38, 31.75) --
	(309.38,211.31);

\path[draw=drawColor,line width= 0.6pt,line join=round] ( 38.56, 36.39) --
	(355.85, 36.39);

\path[draw=drawColor,line width= 0.6pt,line join=round] ( 38.56, 92.17) --
	(355.85, 92.17);

\path[draw=drawColor,line width= 0.6pt,line join=round] ( 38.56,147.96) --
	(355.85,147.96);

\path[draw=drawColor,line width= 0.6pt,line join=round] ( 38.56,203.74) --
	(355.85,203.74);

\path[draw=drawColor,line width= 0.6pt,line join=round] ( 85.03, 31.75) --
	( 85.03,211.31);

\path[draw=drawColor,line width= 0.6pt,line join=round] (149.13, 31.75) --
	(149.13,211.31);

\path[draw=drawColor,line width= 0.6pt,line join=round] (213.23, 31.75) --
	(213.23,211.31);

\path[draw=drawColor,line width= 0.6pt,line join=round] (277.33, 31.75) --
	(277.33,211.31);

\path[draw=drawColor,line width= 0.6pt,line join=round] (341.43, 31.75) --
	(341.43,211.31);
\definecolor{drawColor}{RGB}{0,0,0}

\path[draw=drawColor,line width= 0.6pt,line join=round] ( 52.98, 57.38) --
	( 85.03, 44.49) --
	(149.13, 39.92) --
	(213.23, 56.90) --
	(277.33,111.64) --
	(341.43,203.15);
\definecolor{fillColor}{RGB}{0,0,0}

\path[draw=drawColor,line width= 0.4pt,line join=round,line cap=round,fill=fillColor] ( 52.98, 57.38) circle (  1.96);

\path[draw=drawColor,line width= 0.4pt,line join=round,line cap=round,fill=fillColor] ( 85.03, 44.49) circle (  1.96);

\path[draw=drawColor,line width= 0.4pt,line join=round,line cap=round,fill=fillColor] (149.13, 39.92) circle (  1.96);

\path[draw=drawColor,line width= 0.4pt,line join=round,line cap=round,fill=fillColor] (213.23, 56.90) circle (  1.96);

\path[draw=drawColor,line width= 0.4pt,line join=round,line cap=round,fill=fillColor] (277.33,111.64) circle (  1.96);

\path[draw=drawColor,line width= 0.4pt,line join=round,line cap=round,fill=fillColor] (341.43,203.15) circle (  1.96);
\definecolor{drawColor}{gray}{0.20}

\path[draw=drawColor,line width= 0.6pt,line join=round,line cap=round] ( 38.56, 31.75) rectangle (355.85,211.31);
\end{scope}
\begin{scope}
\path[clip] (  0.00,  0.00) rectangle (361.35,216.81);
\definecolor{drawColor}{gray}{0.30}

\node[text=drawColor,anchor=base east,inner sep=0pt, outer sep=0pt, scale=  0.88] at ( 33.61, 33.36) {0.20};

\node[text=drawColor,anchor=base east,inner sep=0pt, outer sep=0pt, scale=  0.88] at ( 33.61, 89.14) {0.25};

\node[text=drawColor,anchor=base east,inner sep=0pt, outer sep=0pt, scale=  0.88] at ( 33.61,144.93) {0.30};

\node[text=drawColor,anchor=base east,inner sep=0pt, outer sep=0pt, scale=  0.88] at ( 33.61,200.71) {0.35};
\end{scope}
\begin{scope}
\path[clip] (  0.00,  0.00) rectangle (361.35,216.81);
\definecolor{drawColor}{gray}{0.20}

\path[draw=drawColor,line width= 0.6pt,line join=round] ( 35.81, 36.39) --
	( 38.56, 36.39);

\path[draw=drawColor,line width= 0.6pt,line join=round] ( 35.81, 92.17) --
	( 38.56, 92.17);

\path[draw=drawColor,line width= 0.6pt,line join=round] ( 35.81,147.96) --
	( 38.56,147.96);

\path[draw=drawColor,line width= 0.6pt,line join=round] ( 35.81,203.74) --
	( 38.56,203.74);
\end{scope}
\begin{scope}
\path[clip] (  0.00,  0.00) rectangle (361.35,216.81);
\definecolor{drawColor}{gray}{0.20}

\path[draw=drawColor,line width= 0.6pt,line join=round] ( 85.03, 29.00) --
	( 85.03, 31.75);

\path[draw=drawColor,line width= 0.6pt,line join=round] (149.13, 29.00) --
	(149.13, 31.75);

\path[draw=drawColor,line width= 0.6pt,line join=round] (213.23, 29.00) --
	(213.23, 31.75);

\path[draw=drawColor,line width= 0.6pt,line join=round] (277.33, 29.00) --
	(277.33, 31.75);

\path[draw=drawColor,line width= 0.6pt,line join=round] (341.43, 29.00) --
	(341.43, 31.75);
\end{scope}
\begin{scope}
\path[clip] (  0.00,  0.00) rectangle (361.35,216.81);
\definecolor{drawColor}{gray}{0.30}

\node[text=drawColor,anchor=base,inner sep=0pt, outer sep=0pt, scale=  0.88] at ( 85.03, 20.74) {10};

\node[text=drawColor,anchor=base,inner sep=0pt, outer sep=0pt, scale=  0.88] at (149.13, 20.74) {20};

\node[text=drawColor,anchor=base,inner sep=0pt, outer sep=0pt, scale=  0.88] at (213.23, 20.74) {30};

\node[text=drawColor,anchor=base,inner sep=0pt, outer sep=0pt, scale=  0.88] at (277.33, 20.74) {40};

\node[text=drawColor,anchor=base,inner sep=0pt, outer sep=0pt, scale=  0.88] at (341.43, 20.74) {50};
\end{scope}
\begin{scope}
\path[clip] (  0.00,  0.00) rectangle (361.35,216.81);
\definecolor{drawColor}{RGB}{0,0,0}

\node[text=drawColor,anchor=base west,inner sep=0pt, outer sep=0pt, scale=  1.10] at (175.06,  8.71) {Value of $c$};

\end{scope}
\begin{scope}
\path[clip] (  0.00,  0.00) rectangle (361.35,216.81);
\definecolor{drawColor}{RGB}{0,0,0}

\node[text=drawColor,rotate= 90.00,anchor=base,inner sep=0pt, outer sep=0pt, scale=  1.10] at ( 13.08,121.53) {RRMSE};
\end{scope}
\end{tikzpicture}
		\caption{Monte Carlo Relative Root Mean Square Error (RRMSE) of the total estimator of $y_1$ for different values of $c$.}\label{fig:mse:c}
\end{figure}

We first investigate the effect of varying the number of components selected in the bootstrap samples $c$ on the Monte Carlo Relative Root Mean Square Error (RRMSE) of the total estimator. Figure~\ref{fig:mse:c} shows the RRMSE of the total estimator of $y_1$ for different values of $c$. We can see that, as $c$ grows, the RRMSE first decreases until a point where it starts to increase. For this variable of interest, the minimum is attained at around $c = 20$. 

We repeat the same procedure to investigate the effect of varying $\alpha$ on the RRMSE of the total estimator. The results are in Figures~\ref{fig:mse:expn}. The total estimator of $y_1$ has the smallest RRMSE around $\alpha = 1/2$. Moreover, its RRMSE is larger than selecting the components with equal probabilities ($\alpha = 0$) when $\alpha$ is larger than 1. This phenomenon depends on the variable of interest, on the prediction power of the first principal components, and on the choice of $c$. It shows however that giving higher probabilities of being selected to the first principal components is not necessarily a good option.

\begin{figure}[htb!]
\centering
\begin{tikzpicture}[x=1pt,y=1pt]
\definecolor{fillColor}{RGB}{255,255,255}
\path[use as bounding box,fill=fillColor,fill opacity=0.00] (0,0) rectangle (361.35,216.81);
\begin{scope}
\path[clip] (  0.00,  0.00) rectangle (361.35,216.81);
\definecolor{drawColor}{RGB}{255,255,255}
\definecolor{fillColor}{RGB}{255,255,255}

\path[draw=drawColor,line width= 0.6pt,line join=round,line cap=round,fill=fillColor] (  0.00,  0.00) rectangle (361.35,216.81);
\end{scope}
\begin{scope}
\path[clip] ( 38.56, 31.75) rectangle (355.85,211.31);
\definecolor{fillColor}{RGB}{255,255,255}

\path[fill=fillColor] ( 38.56, 31.75) rectangle (355.85,211.31);
\definecolor{drawColor}{gray}{0.92}

\path[draw=drawColor,line width= 0.3pt,line join=round] ( 38.56, 48.29) --
	(355.85, 48.29);

\path[draw=drawColor,line width= 0.3pt,line join=round] ( 38.56,109.39) --
	(355.85,109.39);

\path[draw=drawColor,line width= 0.3pt,line join=round] ( 38.56,170.49) --
	(355.85,170.49);

\path[draw=drawColor,line width= 0.3pt,line join=round] ( 89.03, 31.75) --
	( 89.03,211.31);

\path[draw=drawColor,line width= 0.3pt,line join=round] (161.15, 31.75) --
	(161.15,211.31);

\path[draw=drawColor,line width= 0.3pt,line join=round] (233.26, 31.75) --
	(233.26,211.31);

\path[draw=drawColor,line width= 0.3pt,line join=round] (305.37, 31.75) --
	(305.37,211.31);

\path[draw=drawColor,line width= 0.6pt,line join=round] ( 38.56, 78.84) --
	(355.85, 78.84);

\path[draw=drawColor,line width= 0.6pt,line join=round] ( 38.56,139.94) --
	(355.85,139.94);

\path[draw=drawColor,line width= 0.6pt,line join=round] ( 38.56,201.04) --
	(355.85,201.04);

\path[draw=drawColor,line width= 0.6pt,line join=round] ( 52.98, 31.75) --
	( 52.98,211.31);

\path[draw=drawColor,line width= 0.6pt,line join=round] (125.09, 31.75) --
	(125.09,211.31);

\path[draw=drawColor,line width= 0.6pt,line join=round] (197.20, 31.75) --
	(197.20,211.31);

\path[draw=drawColor,line width= 0.6pt,line join=round] (269.32, 31.75) --
	(269.32,211.31);

\path[draw=drawColor,line width= 0.6pt,line join=round] (341.43, 31.75) --
	(341.43,211.31);
\definecolor{drawColor}{RGB}{0,0,0}

\path[draw=drawColor,line width= 0.6pt,line join=round] ( 52.98, 58.19) --
	( 89.03, 39.92) --
	(125.09, 53.30) --
	(161.15, 94.75) --
	(197.20,111.10) --
	(233.26,127.02) --
	(269.32,143.02) --
	(305.37,174.03) --
	(341.43,203.15);
\definecolor{fillColor}{RGB}{0,0,0}

\path[draw=drawColor,line width= 0.4pt,line join=round,line cap=round,fill=fillColor] ( 52.98, 58.19) circle (  1.96);

\path[draw=drawColor,line width= 0.4pt,line join=round,line cap=round,fill=fillColor] ( 89.03, 39.92) circle (  1.96);

\path[draw=drawColor,line width= 0.4pt,line join=round,line cap=round,fill=fillColor] (125.09, 53.30) circle (  1.96);

\path[draw=drawColor,line width= 0.4pt,line join=round,line cap=round,fill=fillColor] (161.15, 94.75) circle (  1.96);

\path[draw=drawColor,line width= 0.4pt,line join=round,line cap=round,fill=fillColor] (197.20,111.10) circle (  1.96);

\path[draw=drawColor,line width= 0.4pt,line join=round,line cap=round,fill=fillColor] (233.26,127.02) circle (  1.96);

\path[draw=drawColor,line width= 0.4pt,line join=round,line cap=round,fill=fillColor] (269.32,143.02) circle (  1.96);

\path[draw=drawColor,line width= 0.4pt,line join=round,line cap=round,fill=fillColor] (305.37,174.03) circle (  1.96);

\path[draw=drawColor,line width= 0.4pt,line join=round,line cap=round,fill=fillColor] (341.43,203.15) circle (  1.96);
\definecolor{drawColor}{gray}{0.20}

\path[draw=drawColor,line width= 0.6pt,line join=round,line cap=round] ( 38.56, 31.75) rectangle (355.85,211.31);
\end{scope}
\begin{scope}
\path[clip] (  0.00,  0.00) rectangle (361.35,216.81);
\definecolor{drawColor}{gray}{0.30}

\node[text=drawColor,anchor=base east,inner sep=0pt, outer sep=0pt, scale=  0.88] at ( 33.61, 75.81) {0.22};

\node[text=drawColor,anchor=base east,inner sep=0pt, outer sep=0pt, scale=  0.88] at ( 33.61,136.91) {0.24};

\node[text=drawColor,anchor=base east,inner sep=0pt, outer sep=0pt, scale=  0.88] at ( 33.61,198.01) {0.26};
\end{scope}
\begin{scope}
\path[clip] (  0.00,  0.00) rectangle (361.35,216.81);
\definecolor{drawColor}{gray}{0.20}

\path[draw=drawColor,line width= 0.6pt,line join=round] ( 35.81, 78.84) --
	( 38.56, 78.84);

\path[draw=drawColor,line width= 0.6pt,line join=round] ( 35.81,139.94) --
	( 38.56,139.94);

\path[draw=drawColor,line width= 0.6pt,line join=round] ( 35.81,201.04) --
	( 38.56,201.04);
\end{scope}
\begin{scope}
\path[clip] (  0.00,  0.00) rectangle (361.35,216.81);
\definecolor{drawColor}{gray}{0.20}

\path[draw=drawColor,line width= 0.6pt,line join=round] ( 52.98, 29.00) --
	( 52.98, 31.75);

\path[draw=drawColor,line width= 0.6pt,line join=round] (125.09, 29.00) --
	(125.09, 31.75);

\path[draw=drawColor,line width= 0.6pt,line join=round] (197.20, 29.00) --
	(197.20, 31.75);

\path[draw=drawColor,line width= 0.6pt,line join=round] (269.32, 29.00) --
	(269.32, 31.75);

\path[draw=drawColor,line width= 0.6pt,line join=round] (341.43, 29.00) --
	(341.43, 31.75);
\end{scope}
\begin{scope}
\path[clip] (  0.00,  0.00) rectangle (361.35,216.81);
\definecolor{drawColor}{gray}{0.30}

\node[text=drawColor,anchor=base,inner sep=0pt, outer sep=0pt, scale=  0.88] at ( 52.98, 20.74) {0};

\node[text=drawColor,anchor=base,inner sep=0pt, outer sep=0pt, scale=  0.88] at (125.09, 20.74) {1};

\node[text=drawColor,anchor=base,inner sep=0pt, outer sep=0pt, scale=  0.88] at (197.20, 20.74) {2};

\node[text=drawColor,anchor=base,inner sep=0pt, outer sep=0pt, scale=  0.88] at (269.32, 20.74) {3};

\node[text=drawColor,anchor=base,inner sep=0pt, outer sep=0pt, scale=  0.88] at (341.43, 20.74) {4};
\end{scope}
\begin{scope}
\path[clip] (  0.00,  0.00) rectangle (361.35,216.81);
\definecolor{drawColor}{RGB}{0,0,0}

\node[text=drawColor,anchor=base west,inner sep=0pt, outer sep=0pt, scale=  1.10] at (174.75,  8.71) {Value of $\alpha$};
\end{scope}
\begin{scope}
\path[clip] (  0.00,  0.00) rectangle (361.35,216.81);
\definecolor{drawColor}{RGB}{0,0,0}

\node[text=drawColor,rotate= 90.00,anchor=base,inner sep=0pt, outer sep=0pt, scale=  1.10] at ( 13.08,121.53) {RRMSE};
\end{scope}
\end{tikzpicture}

\caption{Monte Carlo Relative Root Mean Square Error (RRMSE) of the total estimator of $y_1$ for different values of $\alpha$.}\label{fig:mse:expn}
\end{figure}

\begin{figure}[htb!]
		\centering
\includegraphics[width=1\textwidth]{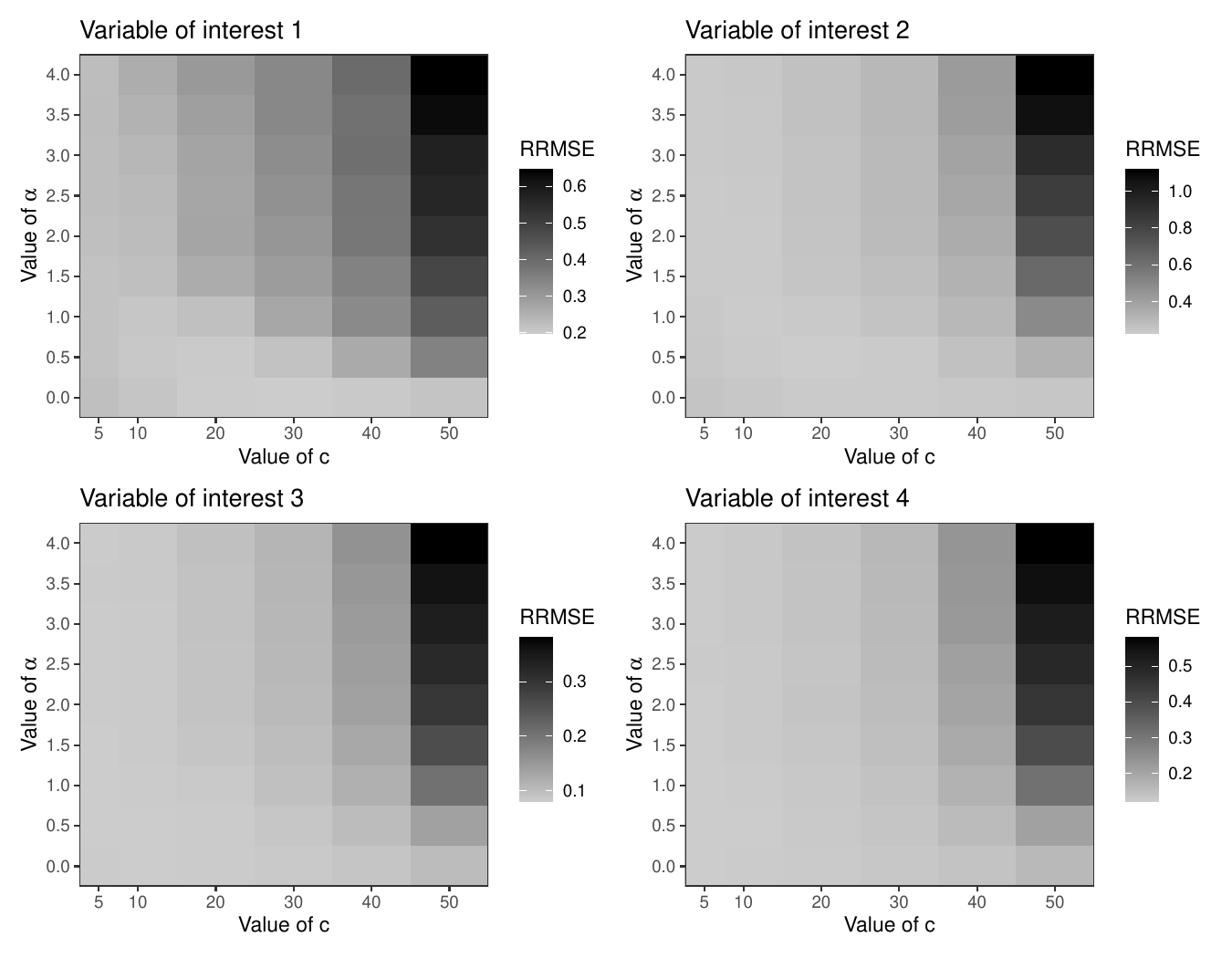}
\caption{Monte Carlo Relative Root Mean Square Error (RRMSE) for four variables of interest $y_1$ to $y_4$ and different values of $\alpha$ and $c$.}\label{fig:rrmse:contour}
\end{figure}

To study the behavior of the proposed estimator when $c$ and $\alpha$ are varied simultaneously, we construct contour plots of the RRMSE for different combinations of these parameters and for each variable of interest. The results are shown in Figure~\ref{fig:rrmse:contour}. Overall, the plots indicate that the RRMSE tends to increase as $c$ and $\alpha$ increase. They also show that the choice $\alpha=0.5$ and $c=10\approx\sqrt{n}=9.3$ provides a good balance across the variables of interest. Although some combinations of $c$ and $\alpha$ yield a lower RRMSE for individual variables, these improvements are not consistent across all variables of interest.

For every simulation run, we compute the minimum and the maximum of the coefficients $g_k$. Then we compute the mean over the 10,000 simulations of the minimum and maximum weights. We also compute the minimum and maximum weight over all 10,000 simulations. We repeat the procedure for different values of $c$ and {$\alpha$. The results are in Figures~\ref{fig:gc} and \ref{fig:galpha}}. We see that, as more components are added to the ensemble members (and therefore to the calibration), the coefficients $g_k$ become more and more dispersed.

\begin{figure}[htb!]
		\centering
\begin{tikzpicture}[x=1pt,y=1pt]
\definecolor{fillColor}{RGB}{255,255,255}
\path[use as bounding box,fill=fillColor,fill opacity=0.00] (0,0) rectangle (361.35,216.81);
\begin{scope}
\path[clip] (  0.00,  0.00) rectangle (361.35,216.81);
\definecolor{drawColor}{RGB}{255,255,255}
\definecolor{fillColor}{RGB}{255,255,255}

\path[draw=drawColor,line width= 0.6pt,line join=round,line cap=round,fill=fillColor] (  0.00,  0.00) rectangle (361.35,216.81);
\end{scope}
\begin{scope}
\path[clip] ( 37.30, 31.75) rectangle (355.85,211.31);
\definecolor{fillColor}{RGB}{255,255,255}

\path[fill=fillColor] ( 37.30, 31.75) rectangle (355.85,211.31);
\definecolor{drawColor}{gray}{0.92}

\path[draw=drawColor,line width= 0.3pt,line join=round] ( 37.30, 57.22) --
	(355.85, 57.22);

\path[draw=drawColor,line width= 0.3pt,line join=round] ( 37.30,105.07) --
	(355.85,105.07);

\path[draw=drawColor,line width= 0.3pt,line join=round] ( 37.30,152.93) --
	(355.85,152.93);

\path[draw=drawColor,line width= 0.3pt,line join=round] ( 37.30,200.78) --
	(355.85,200.78);

\path[draw=drawColor,line width= 0.3pt,line join=round] ( 51.78, 31.75) --
	( 51.78,211.31);

\path[draw=drawColor,line width= 0.3pt,line join=round] (116.13, 31.75) --
	(116.13,211.31);

\path[draw=drawColor,line width= 0.3pt,line join=round] (180.48, 31.75) --
	(180.48,211.31);

\path[draw=drawColor,line width= 0.3pt,line join=round] (244.84, 31.75) --
	(244.84,211.31);

\path[draw=drawColor,line width= 0.3pt,line join=round] (309.19, 31.75) --
	(309.19,211.31);

\path[draw=drawColor,line width= 0.6pt,line join=round] ( 37.30, 33.30) --
	(355.85, 33.30);

\path[draw=drawColor,line width= 0.6pt,line join=round] ( 37.30, 81.15) --
	(355.85, 81.15);

\path[draw=drawColor,line width= 0.6pt,line join=round] ( 37.30,129.00) --
	(355.85,129.00);

\path[draw=drawColor,line width= 0.6pt,line join=round] ( 37.30,176.85) --
	(355.85,176.85);

\path[draw=drawColor,line width= 0.6pt,line join=round] ( 83.95, 31.75) --
	( 83.95,211.31);

\path[draw=drawColor,line width= 0.6pt,line join=round] (148.31, 31.75) --
	(148.31,211.31);

\path[draw=drawColor,line width= 0.6pt,line join=round] (212.66, 31.75) --
	(212.66,211.31);

\path[draw=drawColor,line width= 0.6pt,line join=round] (277.02, 31.75) --
	(277.02,211.31);

\path[draw=drawColor,line width= 0.6pt,line join=round] (341.37, 31.75) --
	(341.37,211.31);
\definecolor{drawColor}{RGB}{0,0,0}

\path[draw=drawColor,line width= 0.6pt,dash pattern=on 4pt off 4pt ,line join=round] ( 51.78,135.62) --
	( 83.95,138.49) --
	(148.31,143.16) --
	(212.66,151.87) --
	(277.02,165.17) --
	(341.37,203.15);

\path[draw=drawColor,line width= 0.6pt,line join=round] ( 51.78,133.94) --
	( 83.95,134.09) --
	(148.31,134.64) --
	(212.66,135.73) --
	(277.02,137.53) --
	(341.37,139.88);

\path[draw=drawColor,line width= 0.6pt,dash pattern=on 4pt off 4pt ,line join=round] ( 51.78,131.53) --
	( 83.95,130.04) --
	(148.31,127.70) --
	(212.66,123.15) --
	(277.02,114.80) --
	(341.37, 39.92);

\path[draw=drawColor,line width= 0.6pt,line join=round] ( 51.78,132.75) --
	( 83.95,132.01) --
	(148.31,131.07) --
	(212.66,130.60) --
	(277.02,130.30) --
	(341.37,128.66);
\definecolor{fillColor}{RGB}{0,0,0}

\path[draw=drawColor,line width= 0.4pt,line join=round,line cap=round,fill=fillColor] ( 51.78,132.75) circle (  1.96);

\path[draw=drawColor,line width= 0.4pt,line join=round,line cap=round,fill=fillColor] ( 51.78,133.94) circle (  1.96);

\path[draw=drawColor,line width= 0.4pt,line join=round,line cap=round,fill=fillColor] ( 51.78,131.53) circle (  1.96);

\path[draw=drawColor,line width= 0.4pt,line join=round,line cap=round,fill=fillColor] ( 51.78,135.62) circle (  1.96);

\path[draw=drawColor,line width= 0.4pt,line join=round,line cap=round,fill=fillColor] ( 83.95,132.01) circle (  1.96);

\path[draw=drawColor,line width= 0.4pt,line join=round,line cap=round,fill=fillColor] ( 83.95,134.09) circle (  1.96);

\path[draw=drawColor,line width= 0.4pt,line join=round,line cap=round,fill=fillColor] ( 83.95,130.04) circle (  1.96);

\path[draw=drawColor,line width= 0.4pt,line join=round,line cap=round,fill=fillColor] ( 83.95,138.49) circle (  1.96);

\path[draw=drawColor,line width= 0.4pt,line join=round,line cap=round,fill=fillColor] (148.31,131.07) circle (  1.96);

\path[draw=drawColor,line width= 0.4pt,line join=round,line cap=round,fill=fillColor] (148.31,134.64) circle (  1.96);

\path[draw=drawColor,line width= 0.4pt,line join=round,line cap=round,fill=fillColor] (148.31,127.70) circle (  1.96);

\path[draw=drawColor,line width= 0.4pt,line join=round,line cap=round,fill=fillColor] (148.31,143.16) circle (  1.96);

\path[draw=drawColor,line width= 0.4pt,line join=round,line cap=round,fill=fillColor] (212.66,130.60) circle (  1.96);

\path[draw=drawColor,line width= 0.4pt,line join=round,line cap=round,fill=fillColor] (212.66,135.73) circle (  1.96);

\path[draw=drawColor,line width= 0.4pt,line join=round,line cap=round,fill=fillColor] (212.66,123.15) circle (  1.96);

\path[draw=drawColor,line width= 0.4pt,line join=round,line cap=round,fill=fillColor] (212.66,151.87) circle (  1.96);

\path[draw=drawColor,line width= 0.4pt,line join=round,line cap=round,fill=fillColor] (277.02,130.30) circle (  1.96);

\path[draw=drawColor,line width= 0.4pt,line join=round,line cap=round,fill=fillColor] (277.02,137.53) circle (  1.96);

\path[draw=drawColor,line width= 0.4pt,line join=round,line cap=round,fill=fillColor] (277.02,114.80) circle (  1.96);

\path[draw=drawColor,line width= 0.4pt,line join=round,line cap=round,fill=fillColor] (277.02,165.17) circle (  1.96);

\path[draw=drawColor,line width= 0.4pt,line join=round,line cap=round,fill=fillColor] (341.37,128.66) circle (  1.96);

\path[draw=drawColor,line width= 0.4pt,line join=round,line cap=round,fill=fillColor] (341.37,139.88) circle (  1.96);

\path[draw=drawColor,line width= 0.4pt,line join=round,line cap=round,fill=fillColor] (341.37, 39.92) circle (  1.96);

\path[draw=drawColor,line width= 0.4pt,line join=round,line cap=round,fill=fillColor] (341.37,203.15) circle (  1.96);
\definecolor{drawColor}{gray}{0.20}

\path[draw=drawColor,line width= 0.6pt,line join=round,line cap=round] ( 37.30, 31.75) rectangle (355.85,211.31);
\end{scope}
\begin{scope}
\path[clip] (  0.00,  0.00) rectangle (361.35,216.81);
\definecolor{drawColor}{gray}{0.30}

\node[text=drawColor,anchor=base east,inner sep=0pt, outer sep=0pt, scale=  0.88] at ( 32.35, 30.27) {-20};

\node[text=drawColor,anchor=base east,inner sep=0pt, outer sep=0pt, scale=  0.88] at ( 32.35, 78.12) {-10};

\node[text=drawColor,anchor=base east,inner sep=0pt, outer sep=0pt, scale=  0.88] at ( 32.35,125.97) {0};

\node[text=drawColor,anchor=base east,inner sep=0pt, outer sep=0pt, scale=  0.88] at ( 32.35,173.82) {10};
\end{scope}
\begin{scope}
\path[clip] (  0.00,  0.00) rectangle (361.35,216.81);
\definecolor{drawColor}{gray}{0.20}

\path[draw=drawColor,line width= 0.6pt,line join=round] ( 34.55, 33.30) --
	( 37.30, 33.30);

\path[draw=drawColor,line width= 0.6pt,line join=round] ( 34.55, 81.15) --
	( 37.30, 81.15);

\path[draw=drawColor,line width= 0.6pt,line join=round] ( 34.55,129.00) --
	( 37.30,129.00);

\path[draw=drawColor,line width= 0.6pt,line join=round] ( 34.55,176.85) --
	( 37.30,176.85);
\end{scope}
\begin{scope}
\path[clip] (  0.00,  0.00) rectangle (361.35,216.81);
\definecolor{drawColor}{gray}{0.20}

\path[draw=drawColor,line width= 0.6pt,line join=round] ( 83.95, 29.00) --
	( 83.95, 31.75);

\path[draw=drawColor,line width= 0.6pt,line join=round] (148.31, 29.00) --
	(148.31, 31.75);

\path[draw=drawColor,line width= 0.6pt,line join=round] (212.66, 29.00) --
	(212.66, 31.75);

\path[draw=drawColor,line width= 0.6pt,line join=round] (277.02, 29.00) --
	(277.02, 31.75);

\path[draw=drawColor,line width= 0.6pt,line join=round] (341.37, 29.00) --
	(341.37, 31.75);
\end{scope}
\begin{scope}
\path[clip] (  0.00,  0.00) rectangle (361.35,216.81);
\definecolor{drawColor}{gray}{0.30}

\node[text=drawColor,anchor=base,inner sep=0pt, outer sep=0pt, scale=  0.88] at ( 83.95, 20.74) {10};

\node[text=drawColor,anchor=base,inner sep=0pt, outer sep=0pt, scale=  0.88] at (148.31, 20.74) {20};

\node[text=drawColor,anchor=base,inner sep=0pt, outer sep=0pt, scale=  0.88] at (212.66, 20.74) {30};

\node[text=drawColor,anchor=base,inner sep=0pt, outer sep=0pt, scale=  0.88] at (277.02, 20.74) {40};

\node[text=drawColor,anchor=base,inner sep=0pt, outer sep=0pt, scale=  0.88] at (341.37, 20.74) {50};
\end{scope}
\begin{scope}
\path[clip] (  0.00,  0.00) rectangle (361.35,216.81);
\definecolor{drawColor}{RGB}{0,0,0}

\node[text=drawColor,anchor=base west,inner sep=0pt, outer sep=0pt, scale=  1.10] at (174.43,  8.71) {Value of $c$};
\end{scope}
\begin{scope}
\path[clip] (  0.00,  0.00) rectangle (361.35,216.81);
\definecolor{drawColor}{RGB}{0,0,0}

\node[text=drawColor,rotate= 90.00,anchor=base west,inner sep=0pt, outer sep=0pt, scale=  1.10] at ( 13.08, 89.23) {Coefficients $g_k$};
\end{scope}
\end{tikzpicture}

		\caption{Mean over 10,000 simulations of the minimum and maximum coefficients $g_k$ (solid lines); minimum and maximum coefficients $g_k$ over 10,000 simulations (dashed). Different values of $c$ are considered.}\label{fig:gc}
\end{figure}

We also see that the coefficients $g_k$ become more and more dispersed as $\alpha$ increases. This phenomenon also depends on the variable of interest, on the prediction power of the first principal components, and on the choice of $c$. The general conclusion is that our proposed method can perform well when the parameters $c$ and $\alpha$ are appropriately selected. In particular, the suggested values $c=10$, which is close to $\sqrt{q}=9.3$, and $\alpha=0.5$ provide good performance across the variables of interest.

\begin{figure}[htb!]
		\centering
\begin{tikzpicture}[x=1pt,y=1pt]
\definecolor{fillColor}{RGB}{255,255,255}
\path[use as bounding box,fill=fillColor,fill opacity=0.00] (0,0) rectangle (361.35,216.81);
\begin{scope}
\path[clip] (  0.00,  0.00) rectangle (361.35,216.81);
\definecolor{drawColor}{RGB}{255,255,255}
\definecolor{fillColor}{RGB}{255,255,255}

\path[draw=drawColor,line width= 0.6pt,line join=round,line cap=round,fill=fillColor] (  0.00,  0.00) rectangle (361.35,216.81);
\end{scope}
\begin{scope}
\path[clip] ( 29.96, 31.75) rectangle (355.85,211.31);
\definecolor{fillColor}{RGB}{255,255,255}

\path[fill=fillColor] ( 29.96, 31.75) rectangle (355.85,211.31);
\definecolor{drawColor}{gray}{0.92}

\path[draw=drawColor,line width= 0.3pt,line join=round] ( 29.96, 51.51) --
	(355.85, 51.51);

\path[draw=drawColor,line width= 0.3pt,line join=round] ( 29.96, 96.29) --
	(355.85, 96.29);

\path[draw=drawColor,line width= 0.3pt,line join=round] ( 29.96,141.06) --
	(355.85,141.06);

\path[draw=drawColor,line width= 0.3pt,line join=round] ( 29.96,185.84) --
	(355.85,185.84);

\path[draw=drawColor,line width= 0.3pt,line join=round] ( 81.81, 31.75) --
	( 81.81,211.31);

\path[draw=drawColor,line width= 0.3pt,line join=round] (155.88, 31.75) --
	(155.88,211.31);

\path[draw=drawColor,line width= 0.3pt,line join=round] (229.94, 31.75) --
	(229.94,211.31);

\path[draw=drawColor,line width= 0.3pt,line join=round] (304.00, 31.75) --
	(304.00,211.31);

\path[draw=drawColor,line width= 0.6pt,line join=round] ( 29.96, 73.90) --
	(355.85, 73.90);

\path[draw=drawColor,line width= 0.6pt,line join=round] ( 29.96,118.67) --
	(355.85,118.67);

\path[draw=drawColor,line width= 0.6pt,line join=round] ( 29.96,163.45) --
	(355.85,163.45);

\path[draw=drawColor,line width= 0.6pt,line join=round] ( 29.96,208.23) --
	(355.85,208.23);

\path[draw=drawColor,line width= 0.6pt,line join=round] ( 44.78, 31.75) --
	( 44.78,211.31);

\path[draw=drawColor,line width= 0.6pt,line join=round] (118.84, 31.75) --
	(118.84,211.31);

\path[draw=drawColor,line width= 0.6pt,line join=round] (192.91, 31.75) --
	(192.91,211.31);

\path[draw=drawColor,line width= 0.6pt,line join=round] (266.97, 31.75) --
	(266.97,211.31);

\path[draw=drawColor,line width= 0.6pt,line join=round] (341.04, 31.75) --
	(341.04,211.31);
\definecolor{drawColor}{RGB}{0,0,0}

\path[draw=drawColor,line width= 0.6pt,dash pattern=on 4pt off 4pt ,line join=round] ( 44.78,107.51) --
	( 81.81,118.31) --
	(118.84,143.45) --
	(155.88,147.48) --
	(192.91,162.08) --
	(229.94,191.66) --
	(266.97,203.15) --
	(304.00,197.65) --
	(341.04,199.00);

\path[draw=drawColor,line width= 0.6pt,line join=round] ( 44.78, 98.17) --
	( 81.81, 97.69) --
	(118.84, 98.26) --
	(155.88, 99.35) --
	(192.91, 99.81) --
	(229.94,100.56) --
	(266.97,100.78) --
	(304.00,101.00) --
	(341.04,102.36);

\path[draw=drawColor,line width= 0.6pt,dash pattern=on 4pt off 4pt ,line join=round] ( 44.78, 82.65) --
	( 81.81, 78.79) --
	(118.84, 71.36) --
	(155.88, 59.64) --
	(192.91, 60.36) --
	(229.94, 56.55) --
	(266.97, 45.32) --
	(304.00, 41.28) --
	(341.04, 39.92);

\path[draw=drawColor,line width= 0.6pt,line join=round] ( 44.78, 90.15) --
	( 81.81, 87.98) --
	(118.84, 87.23) --
	(155.88, 85.95) --
	(192.91, 84.90) --
	(229.94, 83.80) --
	(266.97, 83.06) --
	(304.00, 82.20) --
	(341.04, 81.61);
\definecolor{fillColor}{RGB}{0,0,0}

\path[draw=drawColor,line width= 0.4pt,line join=round,line cap=round,fill=fillColor] ( 44.78, 90.15) circle (  1.96);

\path[draw=drawColor,line width= 0.4pt,line join=round,line cap=round,fill=fillColor] ( 44.78, 98.17) circle (  1.96);

\path[draw=drawColor,line width= 0.4pt,line join=round,line cap=round,fill=fillColor] ( 44.78, 82.65) circle (  1.96);

\path[draw=drawColor,line width= 0.4pt,line join=round,line cap=round,fill=fillColor] ( 44.78,107.51) circle (  1.96);

\path[draw=drawColor,line width= 0.4pt,line join=round,line cap=round,fill=fillColor] ( 81.81, 87.98) circle (  1.96);

\path[draw=drawColor,line width= 0.4pt,line join=round,line cap=round,fill=fillColor] ( 81.81, 97.69) circle (  1.96);

\path[draw=drawColor,line width= 0.4pt,line join=round,line cap=round,fill=fillColor] ( 81.81, 78.79) circle (  1.96);

\path[draw=drawColor,line width= 0.4pt,line join=round,line cap=round,fill=fillColor] ( 81.81,118.31) circle (  1.96);

\path[draw=drawColor,line width= 0.4pt,line join=round,line cap=round,fill=fillColor] (118.84, 87.23) circle (  1.96);

\path[draw=drawColor,line width= 0.4pt,line join=round,line cap=round,fill=fillColor] (118.84, 98.26) circle (  1.96);

\path[draw=drawColor,line width= 0.4pt,line join=round,line cap=round,fill=fillColor] (118.84, 71.36) circle (  1.96);

\path[draw=drawColor,line width= 0.4pt,line join=round,line cap=round,fill=fillColor] (118.84,143.45) circle (  1.96);

\path[draw=drawColor,line width= 0.4pt,line join=round,line cap=round,fill=fillColor] (155.88, 85.95) circle (  1.96);

\path[draw=drawColor,line width= 0.4pt,line join=round,line cap=round,fill=fillColor] (155.88, 99.35) circle (  1.96);

\path[draw=drawColor,line width= 0.4pt,line join=round,line cap=round,fill=fillColor] (155.88, 59.64) circle (  1.96);

\path[draw=drawColor,line width= 0.4pt,line join=round,line cap=round,fill=fillColor] (155.88,147.48) circle (  1.96);

\path[draw=drawColor,line width= 0.4pt,line join=round,line cap=round,fill=fillColor] (192.91, 84.90) circle (  1.96);

\path[draw=drawColor,line width= 0.4pt,line join=round,line cap=round,fill=fillColor] (192.91, 99.81) circle (  1.96);

\path[draw=drawColor,line width= 0.4pt,line join=round,line cap=round,fill=fillColor] (192.91, 60.36) circle (  1.96);

\path[draw=drawColor,line width= 0.4pt,line join=round,line cap=round,fill=fillColor] (192.91,162.08) circle (  1.96);

\path[draw=drawColor,line width= 0.4pt,line join=round,line cap=round,fill=fillColor] (229.94, 83.80) circle (  1.96);

\path[draw=drawColor,line width= 0.4pt,line join=round,line cap=round,fill=fillColor] (229.94,100.56) circle (  1.96);

\path[draw=drawColor,line width= 0.4pt,line join=round,line cap=round,fill=fillColor] (229.94, 56.55) circle (  1.96);

\path[draw=drawColor,line width= 0.4pt,line join=round,line cap=round,fill=fillColor] (229.94,191.66) circle (  1.96);

\path[draw=drawColor,line width= 0.4pt,line join=round,line cap=round,fill=fillColor] (266.97, 83.06) circle (  1.96);

\path[draw=drawColor,line width= 0.4pt,line join=round,line cap=round,fill=fillColor] (266.97,100.78) circle (  1.96);

\path[draw=drawColor,line width= 0.4pt,line join=round,line cap=round,fill=fillColor] (266.97, 45.32) circle (  1.96);

\path[draw=drawColor,line width= 0.4pt,line join=round,line cap=round,fill=fillColor] (266.97,203.15) circle (  1.96);

\path[draw=drawColor,line width= 0.4pt,line join=round,line cap=round,fill=fillColor] (304.00, 82.20) circle (  1.96);

\path[draw=drawColor,line width= 0.4pt,line join=round,line cap=round,fill=fillColor] (304.00,101.00) circle (  1.96);

\path[draw=drawColor,line width= 0.4pt,line join=round,line cap=round,fill=fillColor] (304.00, 41.28) circle (  1.96);

\path[draw=drawColor,line width= 0.4pt,line join=round,line cap=round,fill=fillColor] (304.00,197.65) circle (  1.96);

\path[draw=drawColor,line width= 0.4pt,line join=round,line cap=round,fill=fillColor] (341.04, 81.61) circle (  1.96);

\path[draw=drawColor,line width= 0.4pt,line join=round,line cap=round,fill=fillColor] (341.04,102.36) circle (  1.96);

\path[draw=drawColor,line width= 0.4pt,line join=round,line cap=round,fill=fillColor] (341.04, 39.92) circle (  1.96);

\path[draw=drawColor,line width= 0.4pt,line join=round,line cap=round,fill=fillColor] (341.04,199.00) circle (  1.96);
\definecolor{drawColor}{gray}{0.20}

\path[draw=drawColor,line width= 0.6pt,line join=round,line cap=round] ( 29.96, 31.75) rectangle (355.85,211.31);
\end{scope}
\begin{scope}
\path[clip] (  0.00,  0.00) rectangle (361.35,216.81);
\definecolor{drawColor}{gray}{0.30}

\node[text=drawColor,anchor=base east,inner sep=0pt, outer sep=0pt, scale=  0.88] at ( 25.01, 70.87) {0};

\node[text=drawColor,anchor=base east,inner sep=0pt, outer sep=0pt, scale=  0.88] at ( 25.01,115.64) {2};

\node[text=drawColor,anchor=base east,inner sep=0pt, outer sep=0pt, scale=  0.88] at ( 25.01,160.42) {4};

\node[text=drawColor,anchor=base east,inner sep=0pt, outer sep=0pt, scale=  0.88] at ( 25.01,205.19) {6};
\end{scope}
\begin{scope}
\path[clip] (  0.00,  0.00) rectangle (361.35,216.81);
\definecolor{drawColor}{gray}{0.20}

\path[draw=drawColor,line width= 0.6pt,line join=round] ( 27.21, 73.90) --
	( 29.96, 73.90);

\path[draw=drawColor,line width= 0.6pt,line join=round] ( 27.21,118.67) --
	( 29.96,118.67);

\path[draw=drawColor,line width= 0.6pt,line join=round] ( 27.21,163.45) --
	( 29.96,163.45);

\path[draw=drawColor,line width= 0.6pt,line join=round] ( 27.21,208.23) --
	( 29.96,208.23);
\end{scope}
\begin{scope}
\path[clip] (  0.00,  0.00) rectangle (361.35,216.81);
\definecolor{drawColor}{gray}{0.20}

\path[draw=drawColor,line width= 0.6pt,line join=round] ( 44.78, 29.00) --
	( 44.78, 31.75);

\path[draw=drawColor,line width= 0.6pt,line join=round] (118.84, 29.00) --
	(118.84, 31.75);

\path[draw=drawColor,line width= 0.6pt,line join=round] (192.91, 29.00) --
	(192.91, 31.75);

\path[draw=drawColor,line width= 0.6pt,line join=round] (266.97, 29.00) --
	(266.97, 31.75);

\path[draw=drawColor,line width= 0.6pt,line join=round] (341.04, 29.00) --
	(341.04, 31.75);
\end{scope}
\begin{scope}
\path[clip] (  0.00,  0.00) rectangle (361.35,216.81);
\definecolor{drawColor}{gray}{0.30}

\node[text=drawColor,anchor=base,inner sep=0pt, outer sep=0pt, scale=  0.88] at ( 44.78, 20.74) {0};

\node[text=drawColor,anchor=base,inner sep=0pt, outer sep=0pt, scale=  0.88] at (118.84, 20.74) {1};

\node[text=drawColor,anchor=base,inner sep=0pt, outer sep=0pt, scale=  0.88] at (192.91, 20.74) {2};

\node[text=drawColor,anchor=base,inner sep=0pt, outer sep=0pt, scale=  0.88] at (266.97, 20.74) {3};

\node[text=drawColor,anchor=base,inner sep=0pt, outer sep=0pt, scale=  0.88] at (341.04, 20.74) {4};
\end{scope}
\begin{scope}
\path[clip] (  0.00,  0.00) rectangle (361.35,216.81);
\definecolor{drawColor}{RGB}{0,0,0}

\node[text=drawColor,anchor=base west,inner sep=0pt, outer sep=0pt, scale=  1.10] at (170.45,  8.71) {Value of $\alpha$};
\end{scope}
\begin{scope}
\path[clip] (  0.00,  0.00) rectangle (361.35,216.81);
\definecolor{drawColor}{RGB}{0,0,0}

\node[text=drawColor,rotate= 90.00,anchor=base west,inner sep=0pt, outer sep=0pt, scale=  1.10] at ( 13.08, 89.23) {Coefficients $g_k$};

\end{scope}
\end{tikzpicture}

		\caption{Mean over 10,000 simulations of the minimum and maximum coefficients $g_k$ (solid lines), minimum and maximum coefficients $g_k$ (dashed lines) over 10,000 simulations. Different values of $\alpha$ are considered.}\label{fig:galpha}
\end{figure}

\refstepcounter{section}
\section*{Appendix \thesection: Simulation Study on Synthetic Data}\label{sec:sim:synth:data}

We generate a synthetic finite population of size $N=1000$ containing $p=400$ auxiliary variables, of which 300 are continuous and 100 are binary. The continuous auxiliary variables are generated from a standard multivariate normal distribution. The first 150 variables are generated with pairwise correlation $0.3$. The remaining 150 continuous variables are divided into two blocks of 75 variables, with pairwise correlation $0.1$ within each block. Variables belonging to different blocks are uncorrelated.

The binary auxiliary variables are obtained by thresholding one hundred latent standard Gaussian variables. Dependence is introduced in two subsets of these latent variables as follows. For the first subset, consisting of latent variables 1 to 20, the first latent variable $\vb_1$ serves as a common factor and is generated from a standard normal distribution. The remaining latent variables $\vb_2,\ldots,\vb_{20}$ are obtained via
\begin{align}
  \vb_j &= 0.7\cdot \vb_1 + \sqrt{1-0.7^2}\epsb_j, \quad j = 2, \ldots, 20,
\end{align}
where the variables $\epsb_j$ are generated from a standard normal distribution. Similarly, for latent variables 21 to 40, the 21st latent variable $\vb_{21}$ serves as a common factor and is generated from a standard normal distribution. The remaining latent variables $\vb_{22},\ldots,\vb_{40}$ are obtained via
\begin{align}
  \vb_j &= 0.5\cdot \vb_{21} + \sqrt{1-0.5^2}\epsb_j, \quad j = 22, \ldots, 40,
\end{align} 
where the variables $\epsb_j$ are generated from a standard normal distribution. The remaining latent variables 41 to 100 are generated independently from a standard normal distribution. One hundred variable-specific thresholds are randomly generated as follows: for each latent variable, a quantile of a standard normal distribution is selected, with the order generated from a uniform distribution with parameters, minimum and maximum, of 0.2 and 0.8, respectively. The thresholds are used to create the binary auxiliary variables so that a binary variable is one if the value of the latent variable is greater than the corresponding threshold. The marginal proportions of ones in the 100 binary auxiliary variables vary randomly between approximately $0.2$ and $0.8$. The continuous and binary auxiliary variables are then combined and randomly reordered.

Two variables of interest are generated. For $y_1$, a subset of 10 auxiliary variables $x_1, \ldots, x_{10}$ is randomly selected to carry the signal. Variable $y_1$ is generated according to a linear model involving these selected auxiliary variables, with an intercept of 100 and independent Gaussian errors with mean 0 standard deviation 10. The regression coefficients are $3$, $2.5$, $-2$, $1.5$, $-1$, $0.8$, $0.6$, $-0.5$, $0.3$, and $0.2$.

The second variable of interest, $y_2$, is generated from nonlinear transformations $\tilde{x}_1, \ldots, \tilde{x}_{10}$ of the 10 auxiliary variables that carry the signal for the first variable of interest. The first four transformed auxiliary variables are obtained from the product between two consecutive auxiliary variables. That is, we set $\tilde{x}_1 = x_1\cdot x_2$, $\tilde{x}_2 = x_3\cdot x_4$, etc. The last six transformed variables are obtained from the sinus of 2 times the corresponding auxiliary variable. That is, $\tilde{x}_5 = \sin(2x_5)$, $\tilde{x}_6 = \sin(2x_6)$, etc. The second variable of interest is then obtained from the same linear model as for $y_1$, where the transformed auxiliary variables are considered instead of the initial auxiliary variables. This construction provides two different relationships between the variables of interest and the auxiliary information, one linear and one nonlinear with interactions.

The simulations are conducted as described in Section~\ref{sec:sim:dsgn}. To investigate the effects of the number of auxiliary variables $q$ and the number of principal components selected in our proposed method $c$, both quantities are varied simultaneously. We consider the first $q=50$, $200$, and $400$ initial auxiliary variables from the synthetic population and set the number of selected principal components to $c=15$, $30$, and $45$ in PCA and PCE. All combinations of $q$ and $c$ are considered. The results are presented in Tables~\ref{tab:point:syn:y1} and~\ref{tab:point:syn:y2} for $y_1$ and $y_2$, respectively.

These results show that the estimator calibrated on all auxiliary variables (CAL) performs well when $q=50$. This is expected since, in this case, the number of auxiliary variables is small relative to the sample size, with $q=50<n=200$. In such a setting, the usual calibration estimator remains an appropriate choice. In contrast, when the number of auxiliary variables approaches or exceeds the sample size, as for $q=200=n$ and $q=400=2n$, the methods designed for high-dimensional settings (PCA, VE, and PCE) generally improve upon CAL.

In high-dimensional settings (when $q = 200$ or $q=400$), PCE outperforms PCA in terms of variance. For the linear variable of interest $y_1$, the differences between PCE and PCA are relatively small in terms of RB, although PCE performs better than PCA except for $q = 200, c = 30$. Note that the recommendations given in Section~\ref{subsec:para} yield $c = \min(q, \sqrt{n}) = \sqrt{n} \approx 15$ in this setting, so this exception occurs for a value of $c$ larger than the recommended one. The comparison is less clear for the nonlinear variable $y_2$, for which PCA sometimes yields a smaller bias and the differences between the two methods are generally small. However, PCE continues to provide a lower variance.

\begin{table}
\scriptsize
		\centering
		\caption{Synthetic data -- Monte Carlo relative Bias (RB), Relative Standard Deviation (RSD), Relative Root Mean Square Error (RRMSE), and Variance relative to the Monte Variance of the HT estimator (VARrHT) for the variable of interest $y_1$ for five estimators and combinations of the number of auxiliary variables ($q$) and the number of selected components ($c$).}
		\label{tab:point:syn:y1}

\begin{tabular}{rlrrrrrr}
\toprule
$y$ & Method & $q$ & $c$ & RB ($\times 10^3$) & RSD ($\times 10^3$) & RRMSE ($\times 10^3$) & VARrHT\\
\midrule
1 & CAL & 50 & 15 & 0.29 & 7.61 & 7.61 & 1.26\\
1 & PCA & 50 & 15 & -0.03 & 6.83 & 6.83 & 1.01\\
1 & VE & 50 & 15 & 0.07 & 6.66 & 6.66 & 0.97\\
1 & PCE & 50 & 15 & 0.07 & 6.66 & 6.66 & 0.97\\
1 & HT & 50 & 15 & -0.02 & 6.78 & 6.78 & 1.00\\
\addlinespace
1 & CAL & 50 & 30 & 0.39 & 7.58 & 7.59 & 1.22\\
1 & PCA & 50 & 30 & 0.18 & 7.14 & 7.14 & 1.09\\
1 & VE & 50 & 30 & 0.23 & 6.89 & 6.90 & 1.01\\
1 & PCE & 50 & 30 & 0.23 & 6.92 & 6.92 & 1.02\\
1 & HT & 50 & 30 & 0.04 & 6.85 & 6.85 & 1.00\\
\addlinespace
1 & CAL & 50 & 45 & 0.35 & 7.57 & 7.58 & 1.23\\
1 & PCA & 50 & 45 & 0.22 & 7.18 & 7.19 & 1.11\\
1 & VE & 50 & 45 & 0.24 & 6.91 & 6.92 & 1.03\\
1 & PCE & 50 & 45 & 0.31 & 7.36 & 7.37 & 1.16\\
1 & HT & 50 & 45 & 0.10 & 6.82 & 6.82 & 1.00\\
\addlinespace
1 & CAL & 200 & 15 & -1.95 & 189.46 & 189.47 & 775.10\\
1 & PCA & 200 & 15 & -0.09 & 6.80 & 6.80 & 1.00\\
1 & VE & 200 & 15 & -0.03 & 6.63 & 6.63 & 0.95\\
1 & PCE & 200 & 15 & -0.04 & 6.68 & 6.68 & 0.96\\
1 & HT & 200 & 15 & -0.04 & 6.81 & 6.81 & 1.00\\
\addlinespace
1 & CAL & 200 & 30 & -2.12 & 247.94 & 247.95 & 1367.37\\
1 & PCA & 200 & 30 & 0.02 & 7.10 & 7.10 & 1.12\\
1 & VE & 200 & 30 & 0.14 & 6.61 & 6.61 & 0.97\\
1 & PCE & 200 & 30 & 0.14 & 6.60 & 6.61 & 0.97\\
1 & HT & 200 & 30 & 0.12 & 6.71 & 6.71 & 1.00\\
\addlinespace
1 & CAL & 200 & 45 & 0.17 & 203.39 & 203.39 & 905.59\\
1 & PCA & 200 & 45 & -0.19 & 7.44 & 7.45 & 1.21\\
1 & VE & 200 & 45 & -0.03 & 6.74 & 6.74 & 1.00\\
1 & PCE & 200 & 45 & -0.03 & 6.75 & 6.75 & 1.00\\
1 & HT & 200 & 45 & -0.06 & 6.76 & 6.76 & 1.00\\
\addlinespace
1 & CAL & 400 & 15 & -30.89 & 15.89 & 34.74 & 5.69\\
1 & PCA & 400 & 15 & -0.26 & 6.76 & 6.76 & 1.03\\
1 & VE & 400 & 15 & -0.08 & 6.49 & 6.49 & 0.95\\
1 & PCE & 400 & 15 & -0.08 & 6.56 & 6.56 & 0.97\\
1 & HT & 400 & 15 & -0.08 & 6.66 & 6.66 & 1.00\\
\addlinespace
1 & CAL & 400 & 30 & -30.88 & 15.79 & 34.68 & 5.41\\
1 & PCA & 400 & 30 & -0.24 & 7.05 & 7.05 & 1.08\\
1 & VE & 400 & 30 & -0.04 & 6.58 & 6.58 & 0.94\\
1 & PCE & 400 & 30 & -0.05 & 6.62 & 6.62 & 0.95\\
1 & HT & 400 & 30 & -0.07 & 6.79 & 6.79 & 1.00\\
\addlinespace
1 & CAL & 400 & 45 & -30.95 & 15.84 & 34.77 & 5.50\\
1 & PCA & 400 & 45 & -0.14 & 7.33 & 7.34 & 1.18\\
1 & VE & 400 & 45 & -0.02 & 6.58 & 6.58 & 0.95\\
1 & PCE & 400 & 45 & -0.02 & 6.58 & 6.58 & 0.95\\
1 & HT & 400 & 45 & -0.06 & 6.75 & 6.75 & 1.00\\
\bottomrule
\end{tabular}
\end{table}

\begin{table}
\scriptsize
		\centering
		\caption{Synthetic data -- Monte Carlo relative Bias (RB), Relative Standard Deviation (RSD), Relative Root Mean Square Error (RRMSE), and Variance relative to the Monte Variance of the HT estimator (VARrHT) for the variable of interest $y_2$ for five estimators and combinations of the number of auxiliary variables ($q$) and the number of selected components ($c$).}
		\label{tab:point:syn:y2}

\begin{tabular}{rlrrrrrr}
\toprule
$y$ & Method & p & k & RB ($\times 10^3$) & RSD ($\times 10^3$) & RRMSE ($\times 10^3$) & VARrHT\\
\midrule
2 & CAL & 50 & 15 & 0.07 & 8.07 & 8.07 & 1.35\\
2 & PCA & 50 & 15 & -0.04 & 7.26 & 7.26 & 1.09\\
2 & VE & 50 & 15 & 0.00 & 7.05 & 7.05 & 1.03\\
2 & PCE & 50 & 15 & 0.01 & 7.05 & 7.05 & 1.03\\
2 & HT & 50 & 15 & -0.04 & 6.95 & 6.95 & 1.00\\
\addlinespace
2 & CAL & 50 & 30 & 0.22 & 8.20 & 8.20 & 1.35\\
2 & PCA & 50 & 30 & 0.04 & 7.69 & 7.69 & 1.19\\
2 & VE & 50 & 30 & 0.11 & 7.42 & 7.42 & 1.10\\
2 & PCE & 50 & 30 & 0.12 & 7.45 & 7.45 & 1.11\\
2 & HT & 50 & 30 & -0.04 & 7.06 & 7.06 & 1.00\\
\addlinespace
2 & CAL & 50 & 45 & 0.24 & 8.10 & 8.11 & 1.34\\
2 & PCA & 50 & 45 & 0.13 & 7.55 & 7.55 & 1.17\\
2 & VE & 50 & 45 & 0.16 & 7.34 & 7.35 & 1.10\\
2 & PCE & 50 & 45 & 0.23 & 7.85 & 7.86 & 1.26\\
2 & HT & 50 & 45 & 0.05 & 6.99 & 6.99 & 1.00\\
\addlinespace
2 & CAL & 200 & 15 & -0.75 & 192.14 & 192.15 & 762.11\\
2 & PCA & 200 & 15 & 0.18 & 7.22 & 7.22 & 1.08\\
2 & VE & 200 & 15 & 0.09 & 6.99 & 6.99 & 1.01\\
2 & PCE & 200 & 15 & 0.07 & 6.99 & 6.99 & 1.01\\
2 & HT & 200 & 15 & 0.03 & 6.96 & 6.96 & 1.00\\
\addlinespace
2 & CAL & 200 & 30 & 0.97 & 212.96 & 212.96 & 923.06\\
2 & PCA & 200 & 30 & 0.14 & 7.66 & 7.66 & 1.20\\
2 & VE & 200 & 30 & 0.22 & 7.12 & 7.12 & 1.03\\
2 & PCE & 200 & 30 & 0.22 & 7.11 & 7.11 & 1.03\\
2 & HT & 200 & 30 & 0.14 & 7.01 & 7.01 & 1.00\\
\addlinespace
2 & CAL & 200 & 45 & -6.10 & 235.48 & 235.56 & 1148.74\\
2 & PCA & 200 & 45 & -0.13 & 7.97 & 7.97 & 1.32\\
2 & VE & 200 & 45 & -0.03 & 7.18 & 7.18 & 1.07\\
2 & PCE & 200 & 45 & -0.02 & 7.17 & 7.17 & 1.07\\
2 & HT & 200 & 45 & -0.13 & 6.95 & 6.95 & 1.00\\
\addlinespace
2 & CAL & 400 & 15 & -31.50 & 16.85 & 35.73 & 5.85\\
2 & PCA & 400 & 15 & 0.09 & 7.27 & 7.27 & 1.09\\
2 & VE & 400 & 15 & -0.01 & 6.99 & 6.99 & 1.00\\
2 & PCE & 400 & 15 & -0.03 & 6.98 & 6.98 & 1.00\\
2 & HT & 400 & 15 & -0.04 & 6.97 & 6.97 & 1.00\\
\addlinespace
2 & CAL & 400 & 30 & -31.33 & 16.38 & 35.35 & 5.59\\
2 & PCA & 400 & 30 & 0.10 & 7.50 & 7.50 & 1.17\\
2 & VE & 400 & 30 & 0.09 & 6.96 & 6.97 & 1.01\\
2 & PCE & 400 & 30 & 0.08 & 6.95 & 6.95 & 1.00\\
2 & HT & 400 & 30 & 0.03 & 6.93 & 6.93 & 1.00\\
\addlinespace
2 & CAL & 400 & 45 & -31.46 & 16.62 & 35.58 & 5.68\\
2 & PCA & 400 & 45 & 0.09 & 7.97 & 7.97 & 1.30\\
2 & VE & 400 & 45 & 0.10 & 7.09 & 7.09 & 1.03\\
2 & PCE & 400 & 45 & 0.10 & 7.06 & 7.06 & 1.02\\
2 & HT & 400 & 45 & 0.03 & 6.98 & 6.98 & 1.00\\
\bottomrule
\end{tabular}
\end{table}

\bibliographystyle{chicago}

\end{document}